\pdfoutput=1

\documentclass[11pt,twoside,a4paper,cmspaper,final,collab]{cms-tdr}
\def\svnVersion{b0c3e54-D}\def\svnDate{2019/10/19}\def\cmsCernNoTag{CERN-EP-2019-098}\def\cmsCernDate{\today}\def\cmsMessage{"Published in the Journal of High Energy Physics as \href{http://dx.doi.org/10.1007/JHEP11(2019)082}{\doi{10.1007/JHEP11(2019)082}.}"}

\begin{document}\cmsNoteHeader{TOP-17-019}

\hyphenation{had-ron-i-za-tion}
\hyphenation{cal-or-i-me-ter}
\hyphenation{de-vices}
\RCS$Revision$
\RCS$HeadURL$
\RCS$Id$

\newlength\cmsFigWidth
\ifthenelse{\boolean{cms@external}}{\setlength\cmsFigWidth{0.85\columnwidth}}{\setlength\cmsFigWidth{0.4\textwidth}}
\ifthenelse{\boolean{cms@external}}{\providecommand{\cmsLeft}{top\xspace}}{\providecommand{\cmsLeft}{left\xspace}}
\ifthenelse{\boolean{cms@external}}{\providecommand{\cmsRight}{bottom\xspace}}{\providecommand{\cmsRight}{right\xspace}}
\providecommand{\CL}{CL\xspace}
\newcommand{\wjets}{\ensuremath{\PW{+}\text{jets}}\xspace}
\newcommand{\zjets}{\ensuremath{\PZ{+}\text{jets}}\xspace}
\newcommand{\ttbarjets}{\ensuremath{\ttbar{+}\text{jets}}\xspace}
\newcommand{\tttjets}{\ensuremath{\ttbar\PQt{+}\text{jets}}\xspace}
\newcommand{\tttt}{\ensuremath{\cPqt\cPaqt\cPqt\cPaqt}\xspace}
\newcommand{\ttXY}{\ensuremath{\cPqt\cPaqt \text{XY}}\xspace}
\newcommand{\njets}{\ensuremath{N_{\mathrm{j}}}\xspace}
\newcommand{\nltags}{\ensuremath{N_{\text{tags}}^{\text{l}}}\xspace}
\newcommand{\nmtags}{\ensuremath{N_{\text{tags}}^{\text{m}}}\xspace}
\newcommand{\htb}{\ensuremath{\HT^{\text{\PQb}}}\xspace}
\newcommand{\htrat}{\ensuremath{\HT^{\text{ratio}}}\xspace}
\newcommand{\httwom}{\ensuremath{\HT^{\text{2m}}}\xspace}
\newcommand{\thirdjetpt}{\ensuremath{\pt^{\text{j3}}}\xspace}
\newcommand{\fourthjetpt}{\ensuremath{\pt^{\text{j4}}}\xspace}
\newcommand{\leadleppt}{\ensuremath{\pt^{\ell 1}}\xspace}
\newcommand{\eventsph}{\ensuremath{S}\xspace}
\newcommand{\hth}{\ensuremath{C}\xspace}
\newcommand{\drll}{\ensuremath{\Delta R_{\ell\ell}}\xspace}
\newcommand{\drbb}{\ensuremath{\Delta R_{\PQb\PQb}}\xspace}
\newcommand{\leadlepeta}{\ensuremath{\eta^{\ell 1}}\xspace}
\newcommand{\redhadmass}{\ensuremath{M_{\text{red}}^{\text{h}}}\xspace}
\newcommand{\ourLumi}{\ensuremath{35.8}\xspace}
\newcommand{\sigmattttSM}{\ensuremath{\sigma_{\tttt}^{\text{SM}}}\xspace}
\newcommand{\sigmatttt}{\ensuremath{\sigma_{\tttt}}\xspace}
\newcommand{\BDTtrijetone}{\ensuremath{T_{\text{trijet1}}}\xspace}
\newcommand{\BDTtrijettwo}{\ensuremath{T_{\text{trijet2}}}\xspace}
\newcommand{\BDTdilep}{\ensuremath{D_{\tttt}^{\text{DL}}}\xspace}
\newcommand{\BDTljets}{\ensuremath{D_{\tttt}^{\text{SL}}}\xspace}
\newcommand{\htx}{\ensuremath{H_{\mathrm{T}}^{\mathrm{x}}}\xspace}
\newcommand{\irel}{\ensuremath{I_\text{rel}}\xspace}
\newcommand{\opQL}{\ensuremath{\cmsSymbolFace{Q}_\cmsSymbolFace{L}}\xspace}
\newcommand{\opaQL}{\ensuremath{\overline{\cmsSymbolFace{Q}}_\cmsSymbolFace{L}}\xspace}
\newcommand{\WILOR}{\ensuremath{\mathcal{O}^1_{\cPqt\cPqt}}\xspace}
\newcommand{\WILOLONE}{\ensuremath{\mathcal{O}_{\cmsSymbolFace{Q}\cmsSymbolFace{Q}}^{1}}\xspace}
\newcommand{\WILOBONE}{\ensuremath{\mathcal{O}_{\cmsSymbolFace{Q}\cPqt}^{1}}\xspace}
\newcommand{\WILOBEIGHT}{\ensuremath{\mathcal{O}_{\cmsSymbolFace{Q}\cPqt}^{8}}\xspace}
\providecommand{\NA}{\ensuremath{\text{---}}}
\providecommand{\cmsTable}[1]{\resizebox{\textwidth}{!}{#1}}
\newcommand{\T}{\rule{0pt}{2.6ex}}
\newcommand{\B}{\rule[-1.2ex]{0pt}{0pt}}
\newlength\cmsTabSkip\setlength{\cmsTabSkip}{1ex}
\newcommand{\ttttPredThirteen}{\ensuremath{9.2^{+2.9}_{-2.4}}\xspace}

\newcommand{\singlemuontriggerptcut}{\ensuremath{24}\xspace}
\newcommand{\singleelectrontriggerptcut}{\ensuremath{32}\xspace}
\newcommand{\singlemuonleptonptcut}{\ensuremath{26}\xspace}
\newcommand{\singleelectronleptonptcut}{\ensuremath{35}\xspace}
\newcommand{\leptonrelisocut}{\ensuremath{0.15}\xspace}
\newcommand{\leptonrelisocutloose}{\ensuremath{0.25}\xspace}
\newcommand{\mindileptonbtagptcut}{\ensuremath{25}\xspace}
\newcommand{\minjetptcut}{\ensuremath{30}\xspace}
\newcommand{\maxjetetacut}{\ensuremath{2.5}\xspace}
\newcommand{\minhtcut}{\ensuremath{500}\xspace}
\newcommand{\minmetcut}{\ensuremath{50}\xspace}
\newcommand{\xsecmusinglepton}{\ensuremath{10.6}}
\newcommand{\xsecmusingleptonexp}{\ensuremath{9.4}}
\newcommand{\xsecmusingleptonup}{\ensuremath{4.4}}
\newcommand{\xsecmusingleptondown}{\ensuremath{2.9}}
\newcommand{\xsecfbsinglepton}{\ensuremath{97}}
\newcommand{\xsecfbsingleptonexp}{\ensuremath{86}}
\newcommand{\xsecfbsingleptonup}{\ensuremath{40}}
\newcommand{\xsecfbsingleptondown}{\ensuremath{26}}
\newcommand{\xsecmudilepton}{\ensuremath{6.9}}
\newcommand{\xsecmudileptonexp}{\ensuremath{7.3}}
\newcommand{\xsecmudileptonup}{\ensuremath{4.5}}
\newcommand{\xsecmudileptondown}{\ensuremath{2.5}}
\newcommand{\xsecfbdilepton}{\ensuremath{64}}
\newcommand{\xsecfbdileptonexp}{\ensuremath{67}}
\newcommand{\xsecfbdileptonup}{\ensuremath{41}}
\newcommand{\xsecfbdileptondown}{\ensuremath{23}}
\newcommand{\xsecmucombo}{\ensuremath{5.2}}
\newcommand{\xsecmucomboexp}{\ensuremath{5.7}}
\newcommand{\xsecmucomboup}{\ensuremath{2.9}}
\newcommand{\xsecmucombodown}{\ensuremath{1.8}}
\newcommand{\xsecfbcombo}{\ensuremath{48}}
\newcommand{\xsecfbcomboexp}{\ensuremath{52}}
\newcommand{\xsecfbcomboup}{\ensuremath{26}}
\newcommand{\xsecfbcombodown}{\ensuremath{17}}
\newcommand{\XSecMuComboAllobs}{\ensuremath{3.6}}
\newcommand{\XSecMuComboAllexp}{\ensuremath{2.2}}
\newcommand{\XSecMuComboAllup}{\ensuremath{1.1}}
\newcommand{\XSecMuComboAlldown}{\ensuremath{0.7}}
\newcommand{\XSecFbComboAllobs}{\ensuremath{33}}
\newcommand{\XSecFbComboAllexp}{\ensuremath{20}}
\newcommand{\XSecFbComboAllup}{\ensuremath{10}}
\newcommand{\XSecFbComboAlldown}{\ensuremath{6}}
\newcommand{\signiffbsinglepton}{\ensuremath{0.36}}
\newcommand{\signiffbsingleptonexp}{\ensuremath{0.21}}
\newcommand{\xsfbsinglepton}{\ensuremath{15}}
\newcommand{\xsfbsingleptonup}{\ensuremath{42}}
\newcommand{\xsfbsingleptondown}{\ensuremath{15}}
\newcommand{\bestmusinglepton}{\ensuremath{1.6}}
\newcommand{\bestmusingleptonup}{\ensuremath{4.6}}
\newcommand{\bestmusingleptondown}{\ensuremath{1.6}}
\newcommand{\signiffbdilepton}{\ensuremath{0.0}}
\newcommand{\signiffbdileptonexp}{\ensuremath{0.36}}
\newcommand{\bestmudilepton}{\ensuremath{0.0}}
\newcommand{\bestmudileptonup}{\ensuremath{2.7}}
\newcommand{\xsfbdilepton}{\ensuremath{0}}
\newcommand{\xsfbdileptonup}{\ensuremath{25}}
\newcommand{\signiffbss}{\ensuremath{1.6}}
\newcommand{\signiffbssexp}{\ensuremath{1.0}}
\newcommand{\xsfbss}{\ensuremath{17}}
\newcommand{\xsfbssup}{\ensuremath{14}}
\newcommand{\xsfbssdown}{\ensuremath{11}}
\newcommand{\bestmuss}{\ensuremath{1.8}}
\newcommand{\bestmussup}{\ensuremath{1.5}}
\newcommand{\bestmussdown}{\ensuremath{1.2}}
\newcommand{\signiffbcombo}{\ensuremath{0.0}}
\newcommand{\signiffbcomboexp}{\ensuremath{0.40}}
\newcommand{\xsfbcombo}{\ensuremath{0}}
\newcommand{\xsfbcomboup}{\ensuremath{20}}
\newcommand{\bestmucombo}{\ensuremath{0.0}}
\newcommand{\bestmucomboup}{\ensuremath{2.2}}
\newcommand{\signiffbsscombo}{\ensuremath{1.4}}
\newcommand{\signiffbsscomboexp}{\ensuremath{1.1}}
\newcommand{\xsfbsscombounconstrained}{\ensuremath{13}}
\newcommand{\xsfbsscomboupunconstrained}{\ensuremath{11}}
\newcommand{\xsfbsscombodownunconstrained}{\ensuremath{9}}
\newcommand{\bestmusscombounconstrained}{\ensuremath{1.4}}
\newcommand{\bestmusscomboupunconstrained}{\ensuremath{1.2}}
\newcommand{\bestmusscombodownunconstrained}{\ensuremath{1.0}}
\newcommand{\XSecEFTMuComboAllobs}{\ensuremath{3.6}}
\newcommand{\XSecEFTMuComboAllexp}{\ensuremath{3.2}}

\cmsNoteHeader{TOP-17-019}
\title{Search for the production of four top quarks in the single-lepton and opposite-sign dilepton final states in proton-proton collisions at $\sqrt{s}= 13$\TeV}

\date{\today}

\abstract{
A search for the standard model production of four top quarks ($\Pp\Pp\to \cPqt\cPaqt\cPqt\cPaqt$) is reported using single-lepton plus jets and opposite-sign dilepton plus jets signatures. Proton-proton collisions are recorded with the CMS detector at the LHC at a center-of-mass energy of 13\TeV in a sample corresponding to an integrated luminosity of 35.8\fbinv. A multivariate analysis exploiting global event and jet properties is used to discriminate $\cPqt\cPaqt\cPqt\cPaqt$ from $\ttbar$ production. No significant deviation is observed from the predicted background. An upper limit is set on the cross section for $\cPqt\cPaqt\cPqt\cPaqt$ production in the standard model of 48\unit{fb} at 95\% confidence level. When combined with a previous measurement by the CMS experiment from an analysis of other final states, the observed signal significance is 1.4 standard deviations, and the combined cross section measurement is $ 13^{+11}_{-9}$\unit{fb}. The result is also interpreted in the framework of effective field theory. }

\hypersetup{%
pdfauthor={CMS Collaboration},%
pdftitle={Search for the production of four top quarks in the single-lepton and opposite-sign dilepton final states in proton-proton collisions at sqrt(s) = 13 TeV},%
pdfsubject={CMS},%
pdfkeywords={CMS, physics, top quark, rare SM processes, BSM}}

\maketitle

\section{Introduction}
\label{sec:intro}
Many models of physics beyond the standard model (BSM) predict enhanced or modified couplings of top quarks to other particles. This is particularly relevant for processes that have small production cross sections and, therefore, are yet to be observed, such as the production of four top quarks, \tttt. There is considerable interest in the measurement of the \tttt cross section because of its sensitivity to BSM physics, including supersymmetry~\cite{Nilles:1983ge,Martin:1997ns}, composite models~\cite{Cacciapaglia:2015eqa}, top quark compositeness~\cite{Kumar:2009vs}, two-Higgs-doublet models~\cite{Dicus:1994bm,Craig:2015jba,Craig:2016ygr}, and models with extra spatial dimensions~\cite{Cacciapaglia:2009pa,Ducu:2015fda}.
Within the effective field theory (EFT) framework, the contribution of any BSM process to \tttt production can be parameterized in terms of nonrenormalizable effective couplings of the standard model (SM) fields, if the characteristic energy scale, $\Lambda$, of the BSM physics is much larger than the typical energy scale of \tttt production  at the LHC.
A generic interpretation of the \tttt production can be done using the EFT predictions~\cite{Degrande:2010kt}.

\begin{figure}[h!]
\centering
\includegraphics[width=0.25\textwidth]{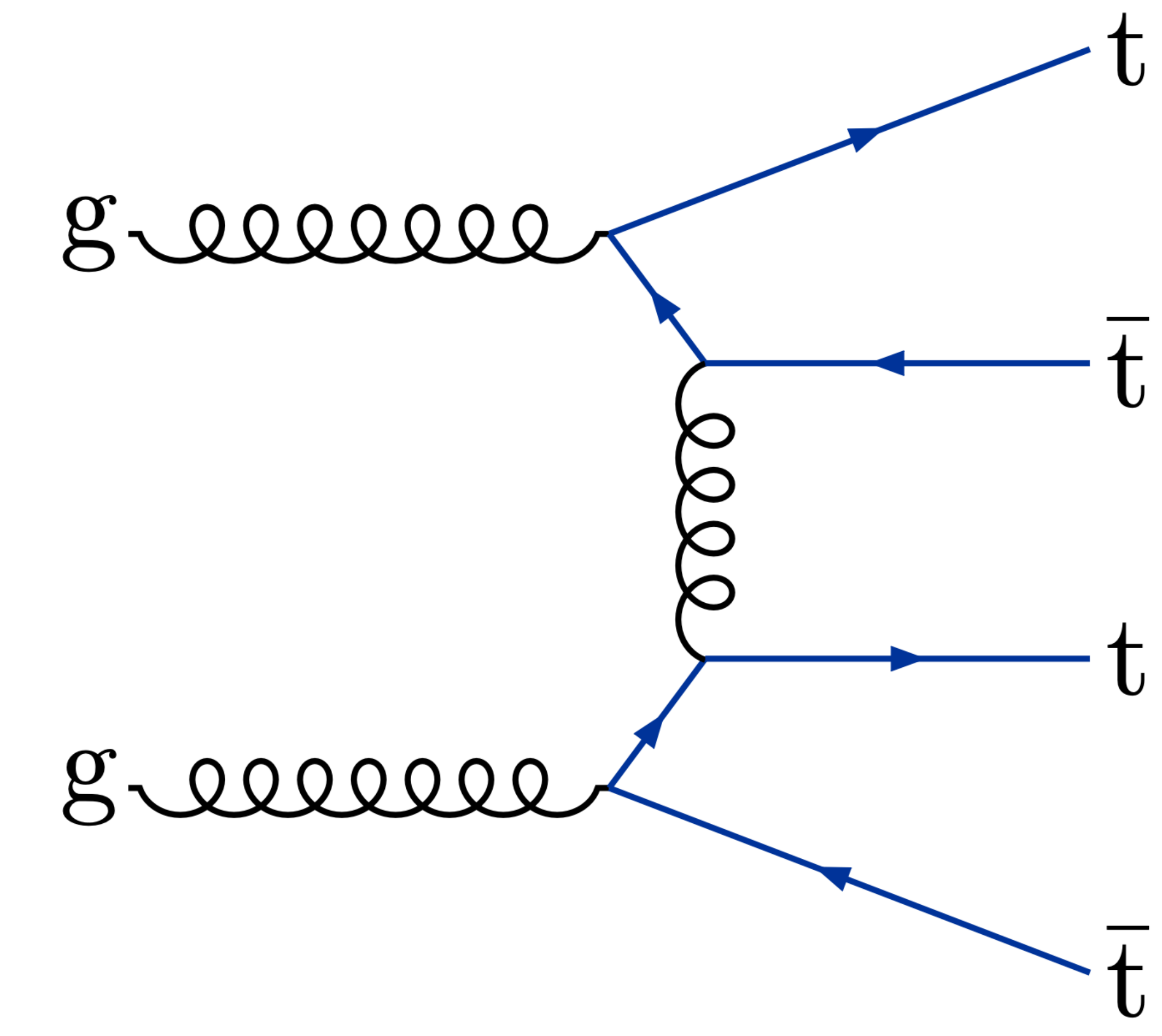} \hspace{40pt}
\includegraphics[width=0.25\textwidth]{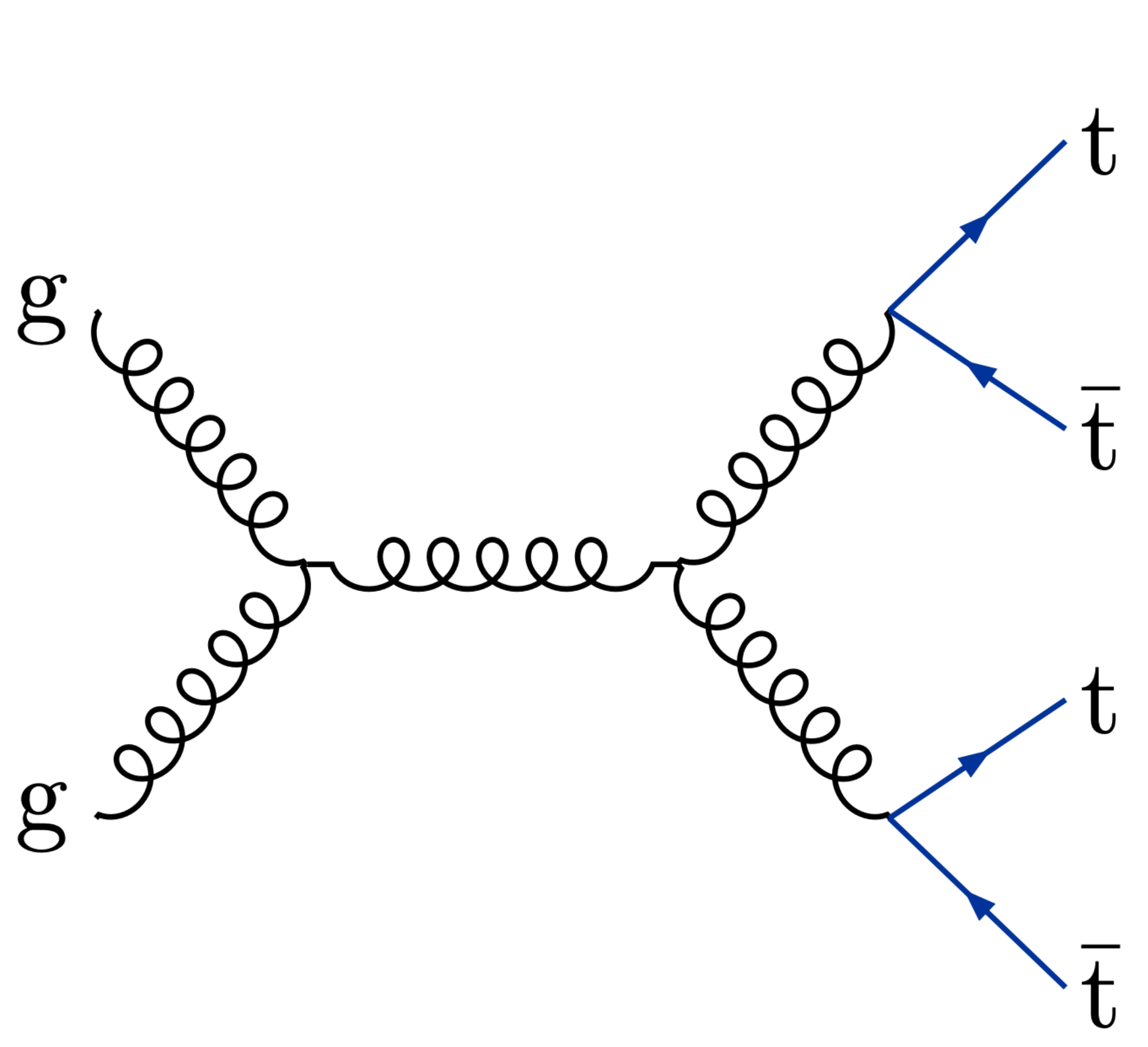}
\caption{Representative Feynman diagrams for $\Pp\Pp\to\tttt$ production at lowest order in the SM.}
\label{fig:feynmantttt}
\end{figure}

The production of four top quarks from proton-proton ($\Pp\Pp$) interactions $\Pp\Pp\to\tttt$ has not yet been observed.
The SM predicts a cross section, at next-to-leading order (NLO), with electroweak corrections (EWK), of $\sigma_{\tttt}^\text{SM}$ of $12.0$\unit{fb} at the center-of-mass energy of 13\TeV~\cite{Frederix:2017wme}. To facilitate comparison with published ATLAS and CMS analyses using comparable data sets, the NLO quantum chromodynamics (QCD) calculation with a value of $\sigma_{\tttt}^\text{SM}=9.2$\unit{fb} is used~\cite{Bevilacqua:2012em,Alwall:2014hca}.
Consequently, the experiments at the CERN LHC may be just approaching sensitivity to the process, provided that it can be separated from the overwhelming background from SM $\ttbar$ events.
The lowest-order Feynman diagrams illustrating typical contributions to SM four top quark production in $\Pp\Pp$ collisions are shown in Fig.~\ref{fig:feynmantttt}.

Searches for \tttt production have been performed at 8\TeV by ATLAS~\cite{Aad:2014pda,Aad:2015kqa,Aad:2015gdg} and CMS\cite{CMStttt8TeV}, and also at 13\TeV (ATLAS (36.1\fbinv\cite{ATLAStttt2018,Aaboud:2018xpj})  and CMS (2.6\fbinv\cite{CMStttt2015})).
The CMS Collaboration measured the \tttt production cross section in a search exploiting same-sign dilepton and multilepton signatures~\cite{TOP-17-009,SUS-16-035} using 13\TeV data (35.9\fbinv) collected in 2016.
The ATLAS Collaboration investigated anomalous $\tttt$ production in events with Lorentz-boosted top quarks identified with top quark tagging techniques~\cite{Aaboud:2018xuw} using 13\TeV data (36.1\fbinv) collected in 2015--2016.

This paper presents a new search in the single-lepton (SL) (\PGm, \Pe)+jets and opposite-sign dilepton (DL) ($\MM$, $\PGmpm \Pemp$, or $\EE$)+jets \tttt decay channels using $\Pp\Pp$ collisions at 13\TeV collected by the CMS experiment in 2016 and corresponding to an integrated luminosity of \ourLumi\fbinv.
For this analysis, only final states containing one or two leptons are
considered, which constitute about 40\% of all \tttt decays.
Compared to the previous analysis~\cite{CMStttt2015}, we have implemented a number of important changes which combine to give a much improved analysis sensitivity. The training process and selection of the input variables for the event-discriminating MVA's (Section~\ref{subsec:mva}) in both the SL and OS dilepton channels has been re-optimized. A new categorization of the signal sensitive regions at large jet and b-tag multiplicities has been introduced, and a revised binning scheme is used to decrease the statistical uncertainties, and improve the signal sensitivity. The categorisation provides additional discrimination against the rare {\ttbar}+boson (H, Z, W, WW/WZ/ZZ) backgrounds. Lastly, a much larger simulated \ttbar data set is used to populate the discriminant bins with high jet multiplicity and high b-tag multiplicity.

\section{The CMS detector}
\label{sec:cms}
The central feature of the CMS apparatus is a superconducting solenoid of 6\unit{m} internal diameter, providing a magnetic field of 3.8\unit{T}.
Within the solenoid volume are a silicon pixel and strip tracker, a lead tungstate crystal electromagnetic calorimeter, and a brass and scintillator hadron calorimeter, each composed of a barrel and two endcap sections.
Forward calorimeters extend the pseudorapidity coverage ($\eta$) provided by the barrel and endcap detectors.
Muons are measured in gas-ionization detectors embedded in the steel flux-return yoke outside the solenoid, in the range $\abs{\eta} < 2.4$.
A more detailed description of the CMS detector, together with a definition of the coordinate system used and the relevant kinematic variables, can be found in Ref.~\cite{Chatrchyan:2008zzk}.

\section{Simulated samples}
\label{sec:mc}
The acceptance for the SM $\Pp\Pp\to\tttt$ process is estimated using samples simulated at NLO precision by the \MGvATNLO 2.2.2 generator~\cite{Alwall:2014hca,Mangano:2006rw}. Only diagrams arising from quantum chromodynamics interactions were taken into account in the simulation. The cross section used to normalize the simulation is the NLO calculation of \ttttPredThirteen\unit{fb}~\cite{Alwall:2014hca}, where the quoted uncertainty incorporates the variation of factorization and renormalization scales used in the calculation of the matrix elements (ME), and the dependence on the choice of parton distribution functions (PDFs). The signal model includes {\sc MadSpin}~\cite{Artoisenet:2012st} and uses the default dynamic scale choice in \MGvATNLO, defined as $\mu_{\mathrm{R,F}}= \frac{1}{2} \Sigma_{\PQt} m_{\mathrm{T}}$.  This is the sum of $m_{\mathrm{T}}$ over each outgoing parton (the four top quarks), divided by two, where $m_{\mathrm{T}} = \sqrt{m^2+\pt^2}$, in which $m$ is the mass of the parton, and \pt is the transverse momentum.

The most important background process is top quark pair production with additional jets (\ttbarjets), that comprises over 90-95\% of the background.
Next in importance are single top (ST) quark processes including $t$-channel and $\PQt\PW$ production. These are followed by \zjets and \wjets electroweak processes (EW), where only the leptonic decay modes of the bosons are considered. Next are rare processes involving the production of a top quark-antiquark pair and a \PZ, \PW, or Higgs~bosons, namely, {\ttbar}+\PZ,\PW,\PH. Finally, \ttbar production in association with dibosons, $\ttbar\PW\PW$, $\ttbar\PW\PZ$, $\ttbar\PZ\PZ$, $\ttbar\PW\PH$, $\ttbar\PZ\PW$, $\ttbar\PH\PH$, and triple top quark production (\tttjets and $\ttbar\PQt\PW$) are considered, processes we collectively denote as \ttXY. Based on their signature resemblance and comparability of production rates to the \tttt signal, {\ttbar}+\PZ and {\ttbar}+\PH are grouped together while {{\ttbar}+\PW} and \ttXY are grouped together in the simulation.

Several Monte Carlo (MC) event generators are used to simulate these processes.
The \ttbarjets process is simulated using the \POWHEG-{\textsc{box}} v2 generator~\cite{Frixione:2007nw,Nason:2004rx,Frixione:2007vw,Alioli:2010xd,Alioli2012} at NLO accuracy for the \ttbar ME, but the \ttbar cross section is normalized to its predicted value at next-to-next-to-leading order (NNLO), which includes soft-gluon corrections, at next-to-next-to-leading-logarithm accuracy, obtained with \textsc{Top++}~2.0~\cite{Czakon:2011xx,ttxsec_1,ttxsec_2,ttxsec_3,ttxsec_4,ttxsec_5,ttxsec_6}. The \POWHEG-{\textsc{box}} simulations are interfaced with \PYTHIA 8.212 using the CUETP8M2T4 tune~\cite{CUETP8M2T4_Tune,Khachatryan:2015pea,Skands:2014pea}. Recent calculations~\cite{ttxsec_6} suggest that next-to-next-to-leading-order effects have an important consequence on the shape of the top quark \PT spectrum that NLO ME generators are unable to reproduce. To allow for this, a parton-level reweighting of the \ttbar simulation has been applied to match the predictions to the data~\cite{Sirunyan:2018ucr,Sirunyan:2018wem}. The correction is applied as a function of the transverse momenta of the parton-level top quark and antiquark after initial- and final-state radiation. Specifically for this result, additional dedicated samples were created that populate the tails in high multiplicity with a factor of 10 more events.

Single top quark $\PQt\PW$ processes are simulated with the \POWHEG-{\textsc{box}} v1 generator~\cite{Re:2010bp}, while $t$-channel processes are simulated with \POWHEG-{\textsc{box}} v2.
Both are interfaced with \PYTHIA 8.212 using the CUETP8M2T4 tune, with the cross sections normalized to the NLO calculations~\cite{stxsec_1,stxsec_2}.
The analysis has been shown~\cite{CMStttt2015} to be insensitive to other ST quark production processes, such as $s$-channel production.

Events with massive gauge bosons and no top quarks (\zjets, \wjets) are simulated using \MGvATNLO~\cite{Alwall:2014hca} at leading-order (LO) accuracy, with up to four additional partons in the ME calculations, and using the MLM matching scheme~\cite{Alwall:2008}. The tune CUETP8M1 is used for the parton shower (PS) and underlying event (UE) modeling. These samples are normalized to their NNLO cross sections~\cite{FEWZ}.

The production of a \ttbar pair in association with a \PW, \PZ and up to one extra parton is simulated using the \MGvATNLO generator at LO accuracy and matched with the PS predictions using the MLM matching scheme. Top quark pair production in association with a Higgs boson, $\ttbar\PH$, is modeled using \POWHEG-{\textsc{box}} v2, interfaced with \PYTHIA 8.212 with the CUETP8M2T4 tune. In this sample, only the dominant $\PH\to\bbbar$ decays are taken into account. These three samples are normalized to the NLO cross sections~\cite{LHCHXSWG4}.
Top quark pair production in association with one or two massive bosons is simulated using the LO ME in the \MGvATNLO generator, and the CUETP8M2T4 tune of  \PYTHIA 8.212 to provide the PS. The cross sections are scaled to their LO values~\cite{LHCHXSWG4}.

For the samples with NLO MEs, the NNPDF3.0NLO~\cite{Ball:2014uwa} PDFs are used, while for LO MEs, the corresponding NNPDF3.0LO PDFs are used. The parton shower, hadronization, and underlying event models implemented in \PYTHIA 8.212~\cite{Sjostrand:2014zea} are used to simulate higher-order processes and nonperturbative aspects of $\Pp\Pp$ collisions.
The NLO simulations use strong coupling constant values of $\alpS(M_{\PZ})=0.137$ and $\alpS(M_{\PZ})=0.113$ for the ME and PS modeling, and the LO simulations use $\alpS(M_{\PZ})=0.130$ for the ME. In all simulations involving the top quark, a mass $m_{\PQt}$ of 172.5\GeV is used.

The \PYTHIA CUETP8M2T4 tune~\cite{CUETP8M2T4_Tune,Khachatryan:2015pea,Skands:2014pea} currently provides the best description of the \ttbar data~\cite{Sirunyan:2018asm,Sirunyan:2018ptc}. The \POWHEG-{\textsc{box}} calculation describes the high-multiplicity tail when this tune is used. The uncertainties cover the differences due to alternative choices of the PS and hadronization models~\cite{CMS-PAS-TOP-16-021}.

All of the simulated samples include an estimate of the additional $\Pp\Pp$ interactions per bunch crossing (pileup), modeled with the \PYTHIA 8.212 program. Corrections are applied to make the simulation of the number of additional interactions representative of that observed in the data. The simulated events are propagated through a simulation of the CMS detector based on \GEANTfour (v.9.4)~\cite{Agostinelli2003250} and reconstructed using the same algorithms as for the collider data.

\section{Data analysis}
\label{sec:method}
\subsection{Event selection}
\label{subsec:eventselection}
The final states considered in this analysis are the single-lepton channel with exactly one muon or electron, (\PGm, \Pe)+jets, and the opposite-sign dilepton channel, ($\MM$, $\PGmpm \Pemp$, $\EE$)+jets.
In all cases, the leptons are expected to originate from the {\PW} bosons arising from top quark decays and thus tend to be isolated, unlike the leptons produced in the decay of unstable hadrons within jets.

Single-lepton events were recorded using a trigger~\cite{Khachatryan:2016bia} that required at least one isolated muon with $\PT > \singlemuontriggerptcut\GeV$ and $\abs{\eta} < 2.4$, or one isolated electron with $\PT > \singleelectrontriggerptcut\GeV$ and $\abs{\eta} < 2.1$.
Dilepton events were recorded using either single-lepton or dilepton triggers. In the case of dilepton triggers, the \PT thresholds for the leading and subleading leptons for the dimuon triggers are 17 and 8\GeV, respectively, 23 and 12\GeV for dielectron triggers, and 23 and 8\GeV for muon-electron triggers, regardless of lepton flavor. Dilepton triggers require $\abs{\eta} < 2.4$ for muons and $\abs{\eta} < 2.5$ for electrons.
The single-lepton triggers were also used in the dilepton channel to increase the efficiency, while retaining the orthogonality of the selections addressing the two final states.

Offline event reconstruction uses the CMS particle-flow (PF) algorithm~\cite{Sirunyan:2017ulk} for particle reconstruction and identification.
Single-lepton events are required to have exactly one isolated muon with $\PT > \singlemuonleptonptcut\GeV$ or one isolated electron with $\PT > \singleelectronleptonptcut\GeV$, either within $\abs{\eta} < 2.1$.
In the dilepton channel, events are required to contain exactly two isolated leptons of opposite sign with $\PT > 25 \GeV$ for the leading and $\PT > 20 \GeV$ for the subleading lepton, within $\abs{\eta} < 2.4$.
Muons must satisfy the criteria described in Ref.~\cite{Sirunyan:2018fpa} and have a relative isolation, $\irel < \leptonrelisocut$.
Electron candidates must satisfy stringent identification criteria, including \irel, which are described in Ref.~\cite{Khachatryan:2015hwa}.
The \irel is defined as the scalar \pt sum of the additional particles consistent with the same vertex as the lepton, within a cone of angular radius $\Delta R = \sqrt{\smash[b]{(\Delta\eta)^{2} + (\Delta\phi)^{2}}}$ = 0.4  around the lepton, divided by the \pt of the lepton, where $\Delta\eta$ and $\Delta\phi$ (in radians) are the differences in pseudorapidity and azimuthal angle, respectively, between the directions of the lepton and the additional particle.
The sum is corrected for the neutral particle contribution from pileup on an event-by-event basis~\cite{Khachatryan:2015hwa, Sirunyan:2018fpa}.
To suppress background events from decays of low-mass resonances and \PZ~bosons, the lepton pairs are required to have an invariant mass greater than 20\GeV and be outside of a 30\GeV window centered on the \PZ~boson mass in both the $\MM$ and $\EE$ channels.
Events containing additional muons with looser relative isolation, $\irel<\leptonrelisocutloose$, or isolated electrons are vetoed.

Each event is required to contain at least one reconstructed vertex. The reconstructed vertex with the largest value of the quadratic sum of the \PT of its associated tracks is considered the primary $\Pp\Pp$ interaction vertex. Jets are reconstructed from the PF candidates using the infrared- and collinear-safe anti-\kt algorithm~\cite{Cacciari:2008gp,Cacciari:2011ma} with a distance parameter of 0.4. Pileup interactions can contribute tracks and calorimetric energy depositions to the jet momentum. To mitigate this effect, charged particles identified as originating from pileup vertices are discarded and the jet is corrected for the remaining contributions~\cite{Cacciari:2008gn,CMS-PAS-JME-14-001}. Jet energy corrections are derived from simulations to bring the measured response of jets to that of particle level jets on average. In situ measurements of the momentum balance in dijet, $\ensuremath{\text{photon}{+}\text{jet}}$, $\ensuremath{\PZ{+}\text{jet}}$, and multijet events are used to account for any residual differences in jet energy scale between real and simulated data~\cite{Khachatryan:2016kdb}. The jet energy resolution is typically 15\% at 10\GeV, 8\% at 100\GeV, and 4\% at 1\TeV. The missing transverse momentum vector \ptvecmiss is computed as the negative vector sum of \PT of all the PF candidates in an event, and its magnitude is denoted as \ptmiss~\cite{CMS-PAS-JME-17-001}. The quantity \ptvecmiss is modified to account for corrections to the energy of the reconstructed jets in the event.

A minimum of seven jets for the single-muon and eight jets for the single-electron channel are required, each of which must have $\PT > \minjetptcut\GeV$ and $\abs{\eta} < \maxjetetacut$. The difference in the jet multiplicity is motivated by the need to reduce the residual contamination from multijet QCD background in the electron channel due to a higher lepton misidentification rate. In the selected events, at least two jets must be tagged as originating from the hadronization of bottom quarks ({\cPqb} jets) using the combined secondary vertex (CSVv2) algorithm at its medium working point~\cite{Sirunyan:2017ezt}. Additional {\cPqb} jet candidates are identified using the CSVv2 algorithm at its loose working point. The two working points, loose and medium, provide different levels of purity and efficiency. The loose working point gives a misidentification rate of approximately 10\% for light-quark and gluon jets, with a {\cPqb} tagging efficiency of about 80\%. The medium working point has a misidentification rate of about 1\% with a {\cPqb} tagging efficiency of about 68\%. The efficiency to tag \cPqc quarks is 12\%. To suppress the small residual QCD background, \ptmiss is required to be larger than \minmetcut\GeV. Studies on the estimation of non-prompt leptons from QCD multijet background by inverting lepton isolation selection criteria have verified that this background is negligible after applying the selection requirements. In addition, a requirement on the scalar sum of the \PT of all jets, $\HT > \minhtcut\GeV$, is applied. The \HT requirement is used to suppress the \ttbar background, while having little effect on the signal acceptance~\cite{CMStttt2015}.

In the dilepton channels, a minimum of four jets is required, each with $\abs{\eta} < 2.4$. Of these, at least two must be {\cPqb}-tagged using the same CSVv2 algorithm with medium working point as was used in the single-lepton channel. While the \PT threshold for non-tagged jets is $\minjetptcut\GeV$ (as for the single-lepton channel), the threshold for {\cPqb}-tagged jets is lowered to $\mindileptonbtagptcut\GeV$ to increase the acceptance for events with multiple {\cPqb} jets.
The $\HT > 500\GeV$ requirement is also applied to the dilepton channels.

\begin{figure}[ht!]
\centering
\includegraphics[width=0.45\textwidth]{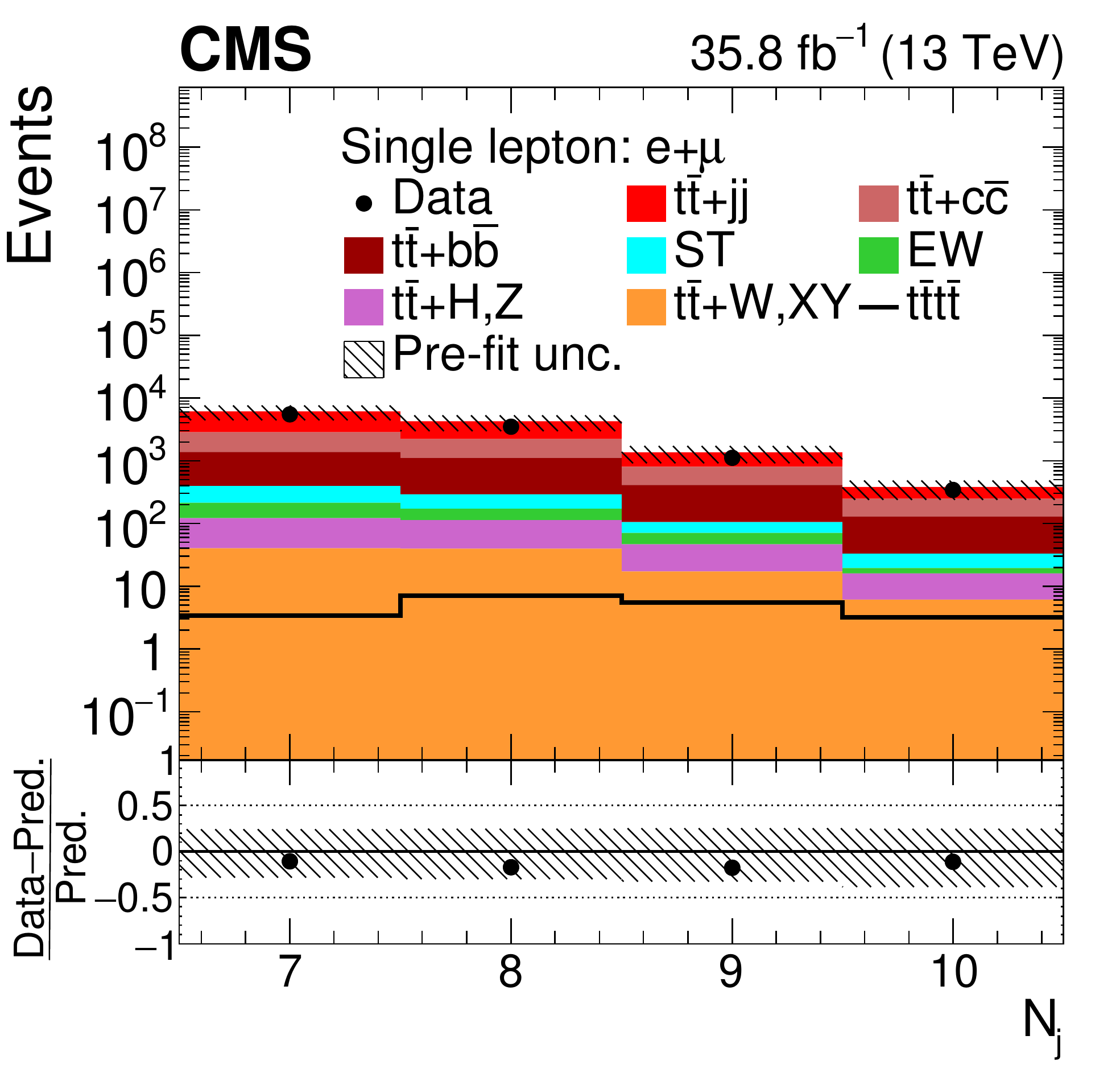}
\includegraphics[width=0.45\textwidth]{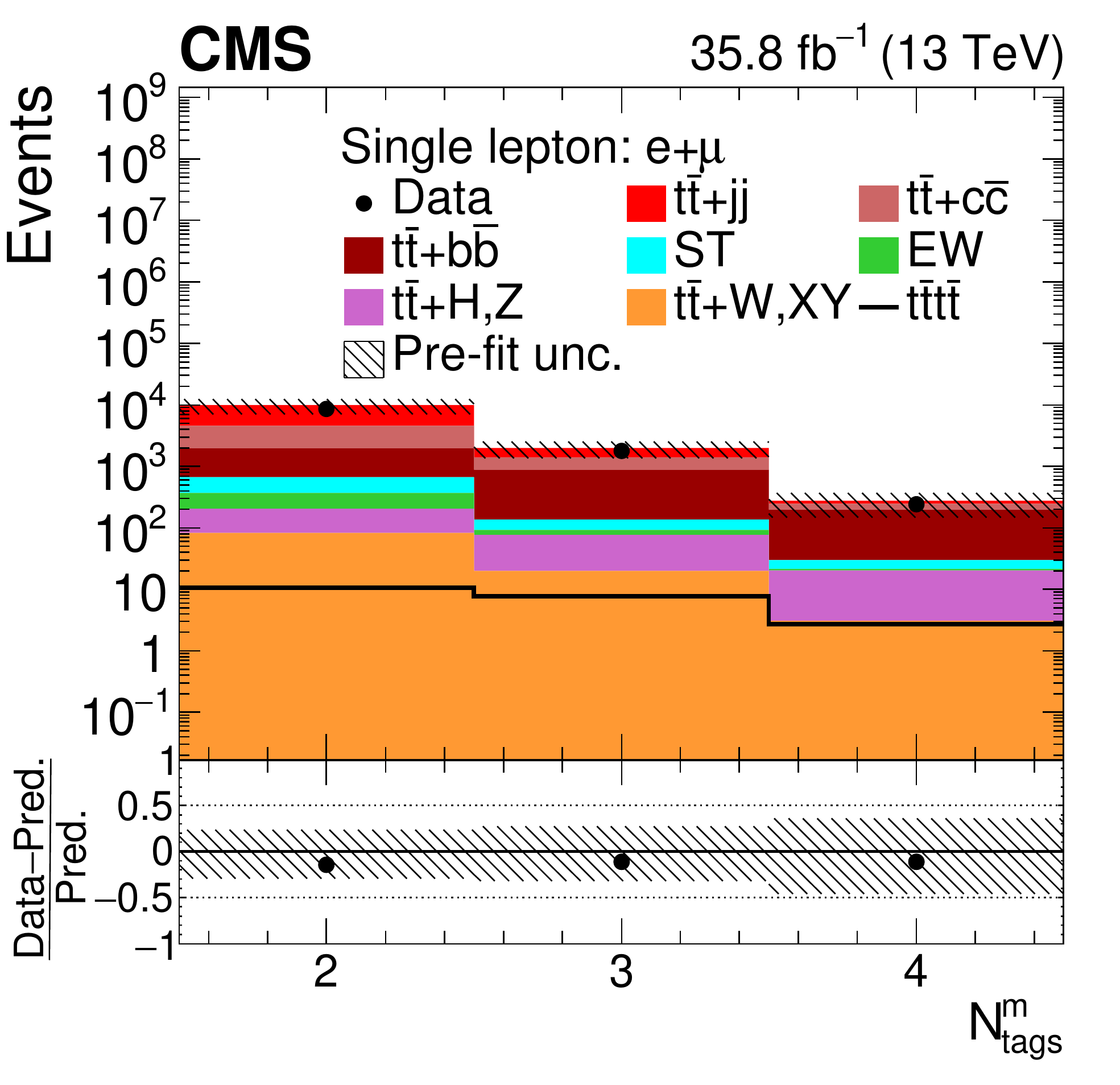}
\\
\includegraphics[width=0.45\textwidth]{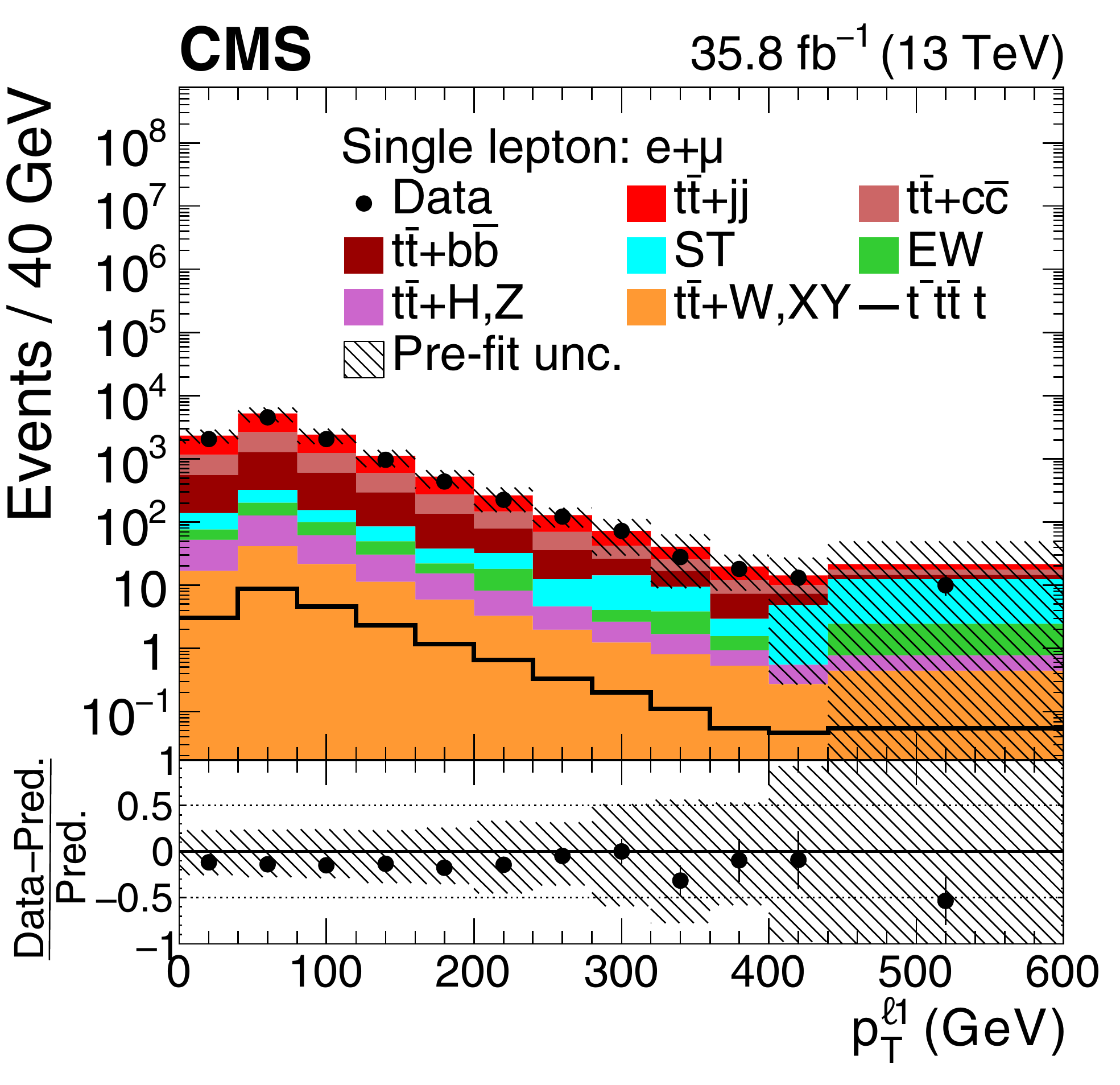}
\includegraphics[width=0.45\textwidth]{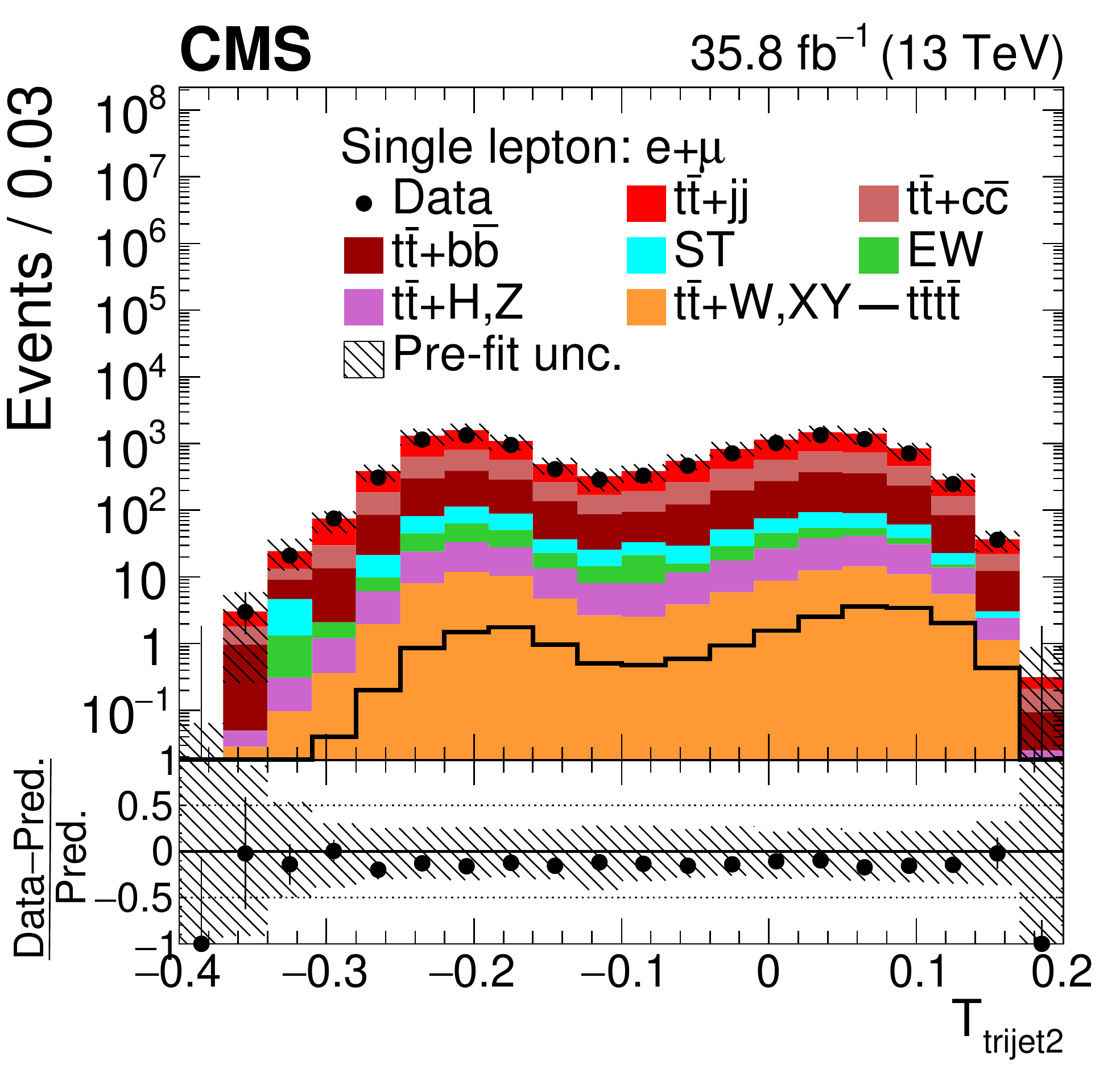}
\caption{Distributions of \njets, \nmtags, \leadleppt and \BDTtrijettwo in the combined single-lepton channels. In the upper panels, the data are shown as dots with error bars representing statistical uncertainties, MC simulations are shown as a histogram. The lower panels show the relative difference between the data and the sum of all of the standard model backgrounds. In each panel, the shaded band represents the total uncertainty in the dominant \ttbar background estimate. See Section~\ref{subsec:mva} for the definitions of the variables. }
\label{fig:controldistributions_SL1}
\end{figure}

\begin{figure}[ht!]
\centering
\includegraphics[width=0.45\textwidth]{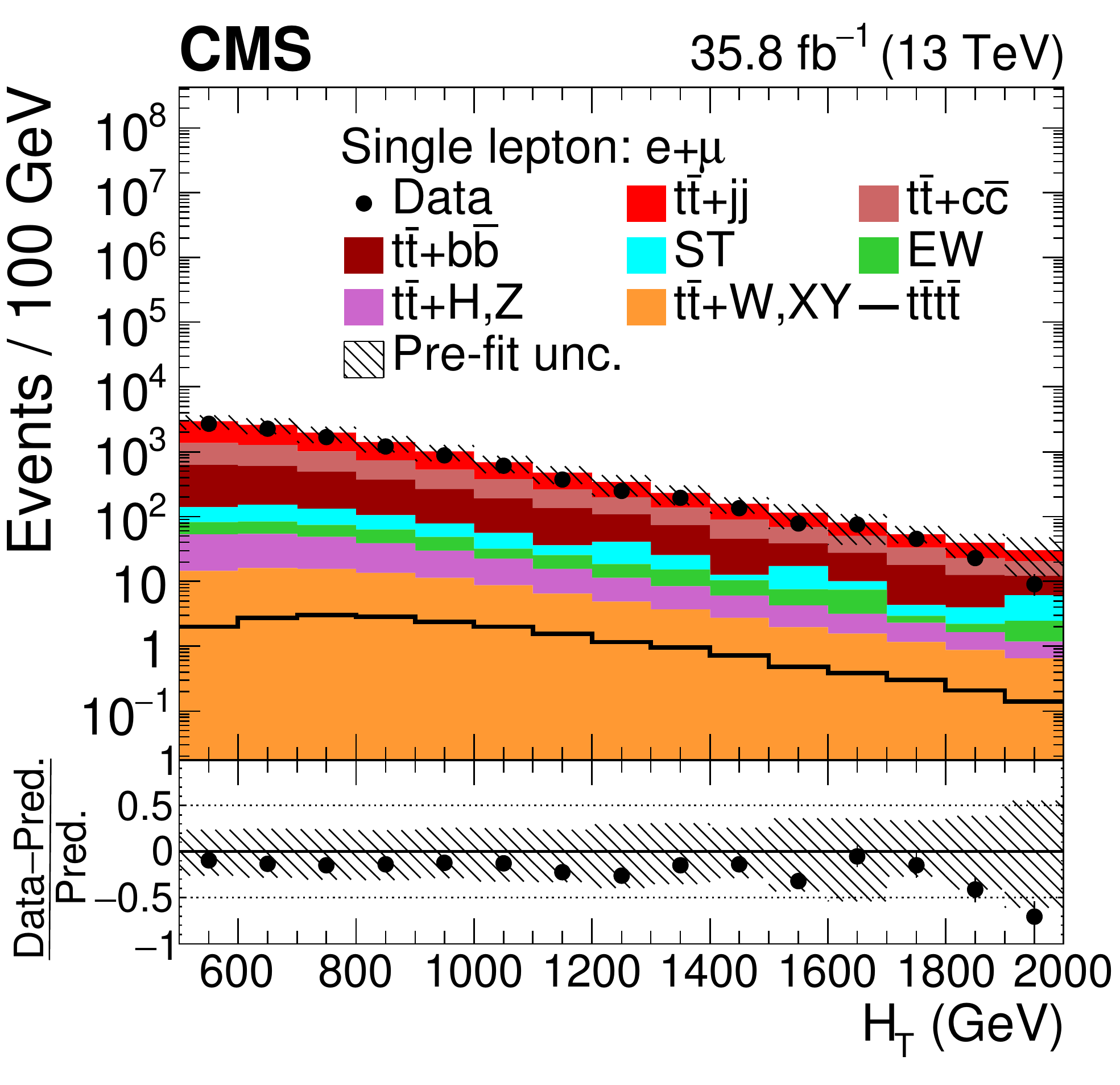}
\includegraphics[width=0.45\textwidth]{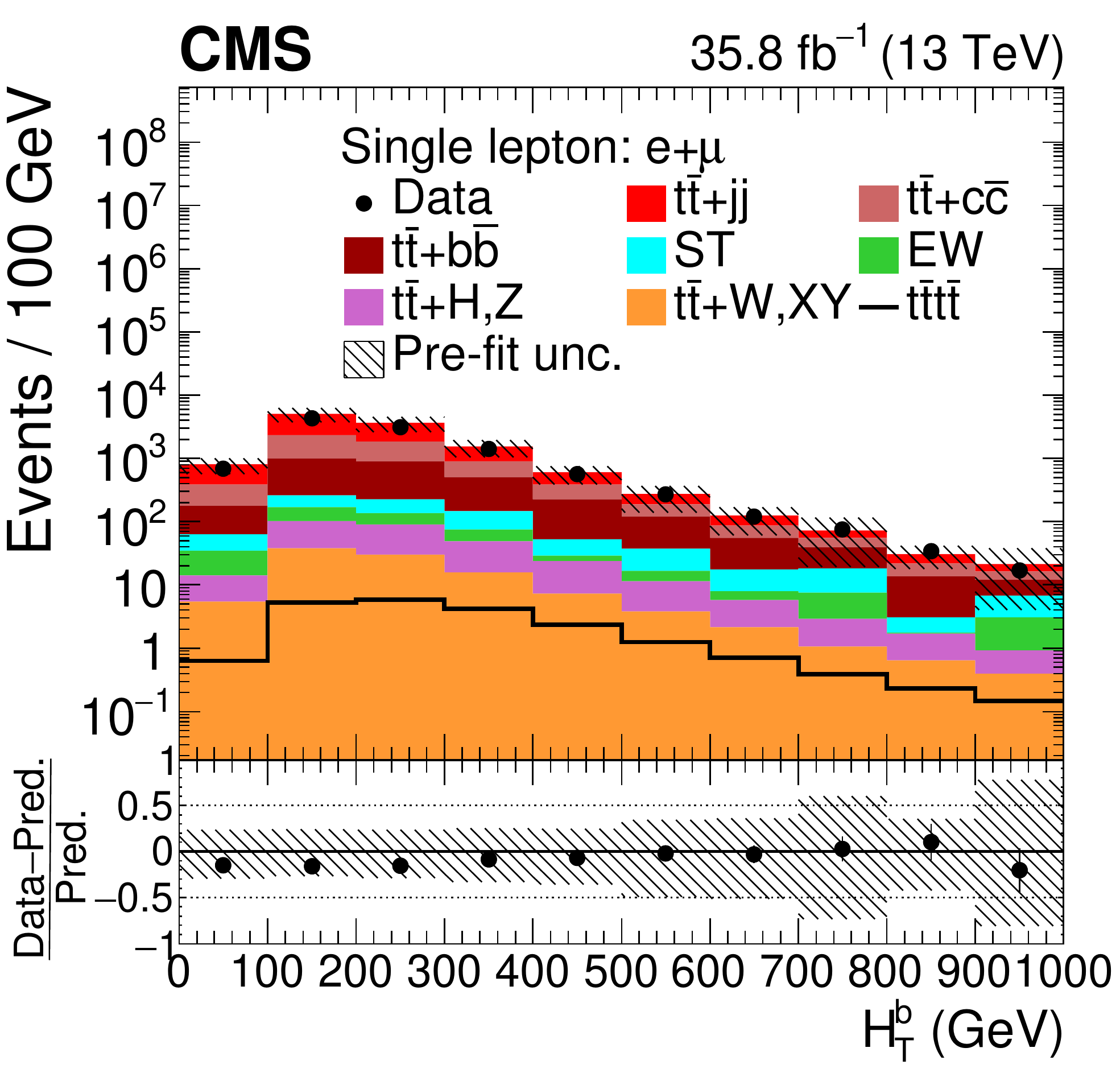}
\\
\includegraphics[width=0.45\textwidth]{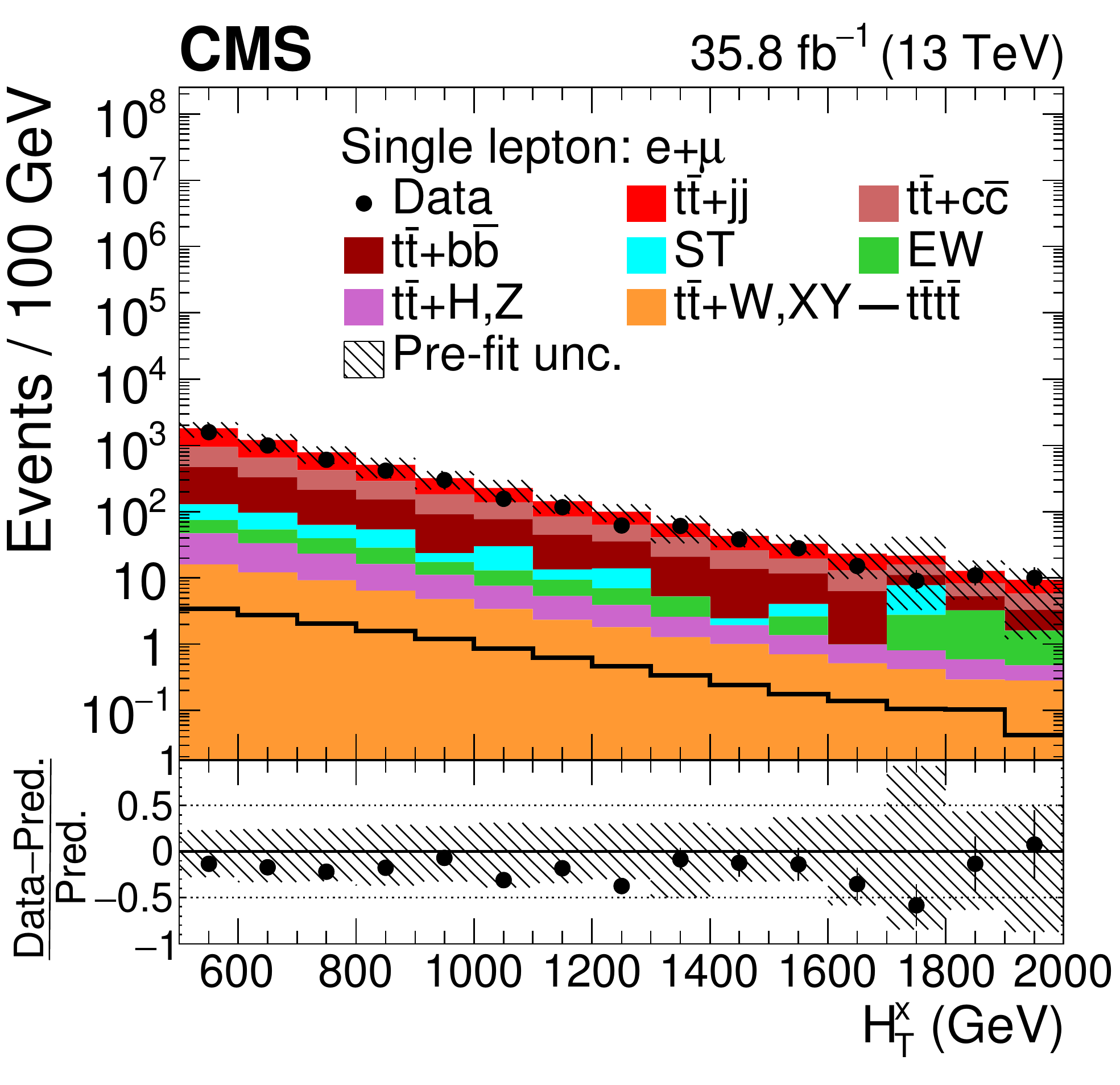}
\includegraphics[width=0.45\textwidth]{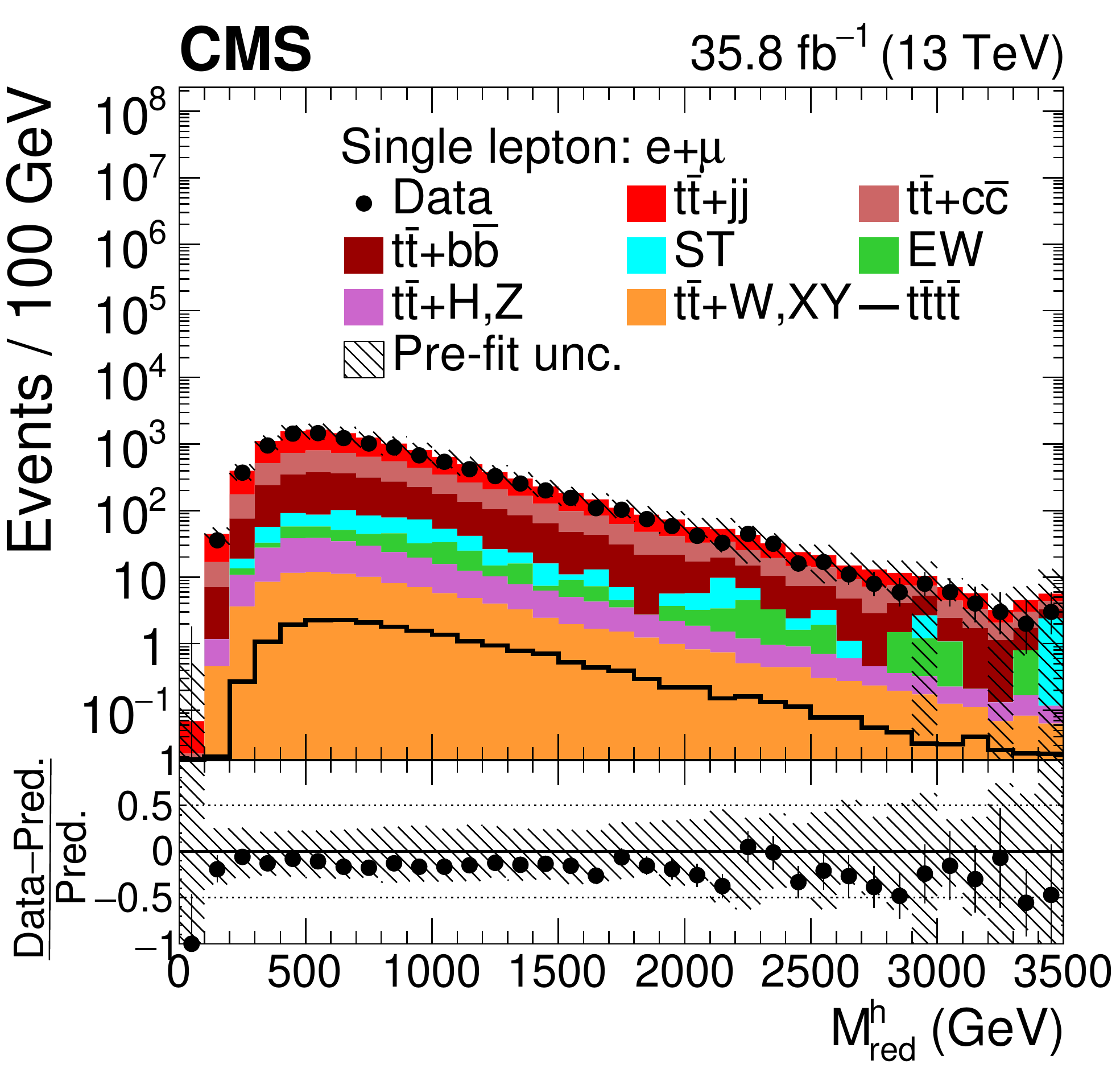}
\caption{Distributions of \HT, \htb, \htx and \redhadmass in the combined single-lepton channel.
In the upper panels, the data are shown as dots with error bars representing statistical uncertainties, MC simulations are shown as a histogram. The lower panels show the relative difference between the data and the sum of all of the standard model backgrounds. In each panel, the shaded band represents the total uncertainty in the dominant \ttbar background estimate. See Section~\ref{subsec:mva} for the definitions of the variables. }
\label{fig:controldistributions_SL2}
\end{figure}

\begin{figure}[ht!]
\centering
\includegraphics[width=0.45\textwidth]{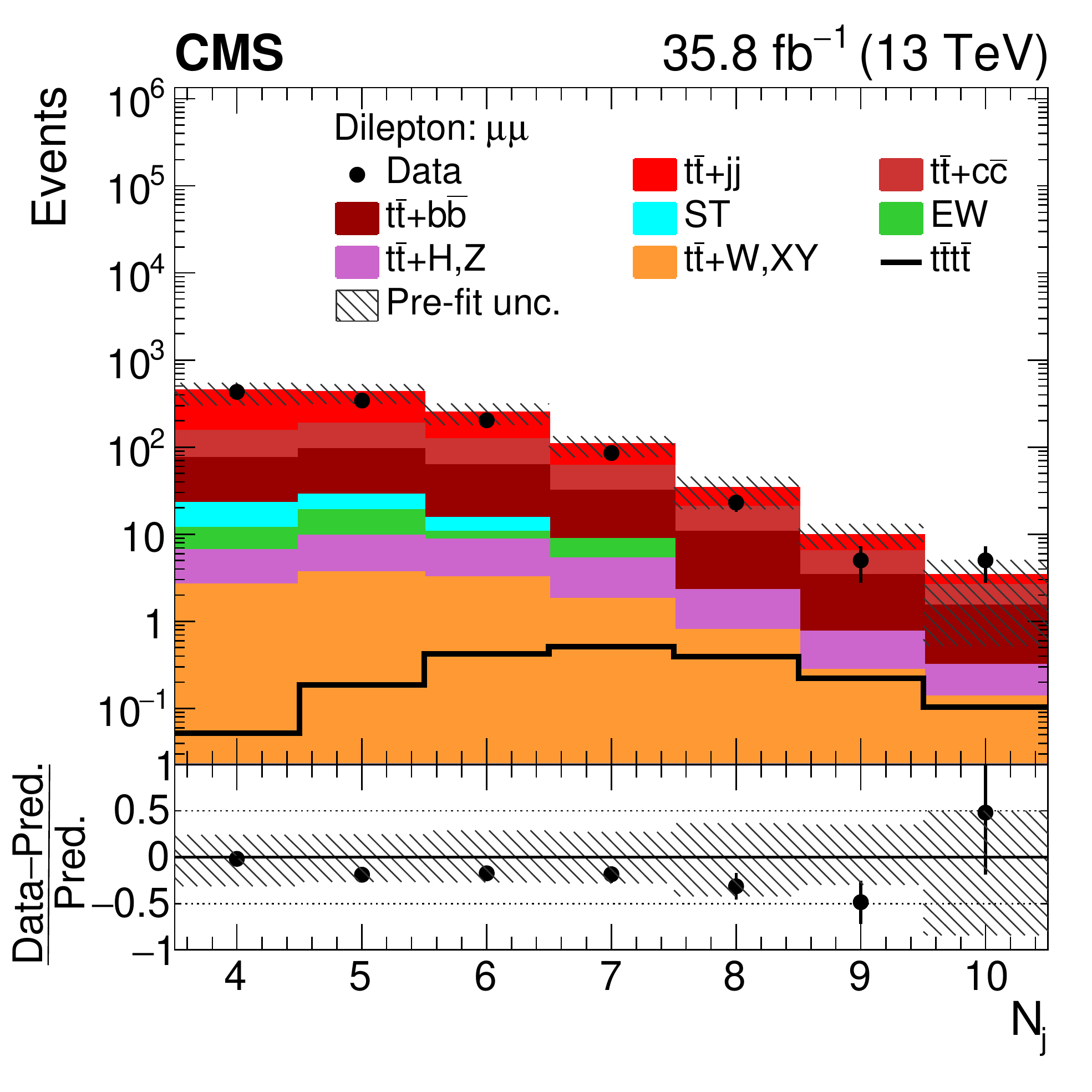}
\includegraphics[width=0.45\textwidth]{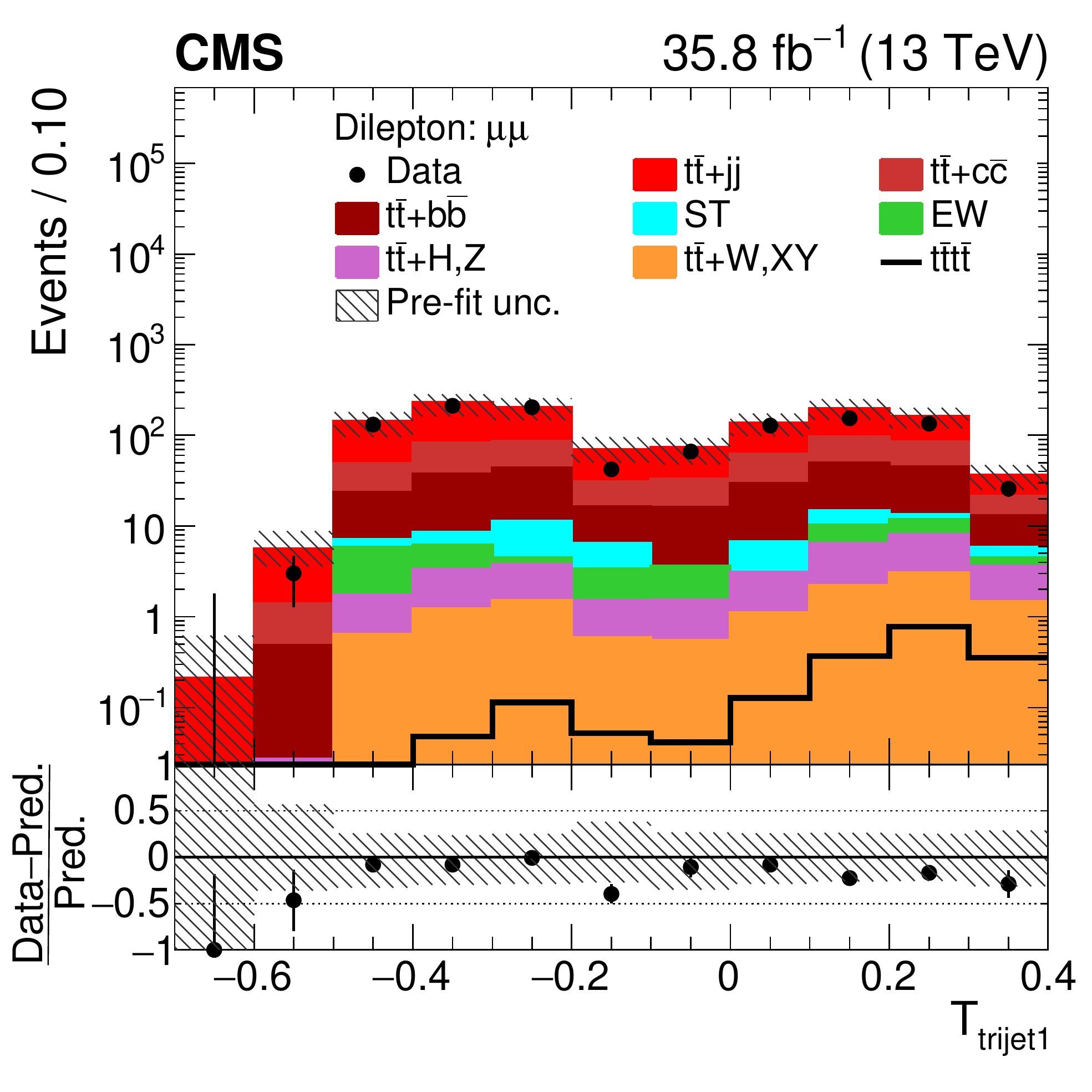}
\\
\includegraphics[width=0.45\textwidth]{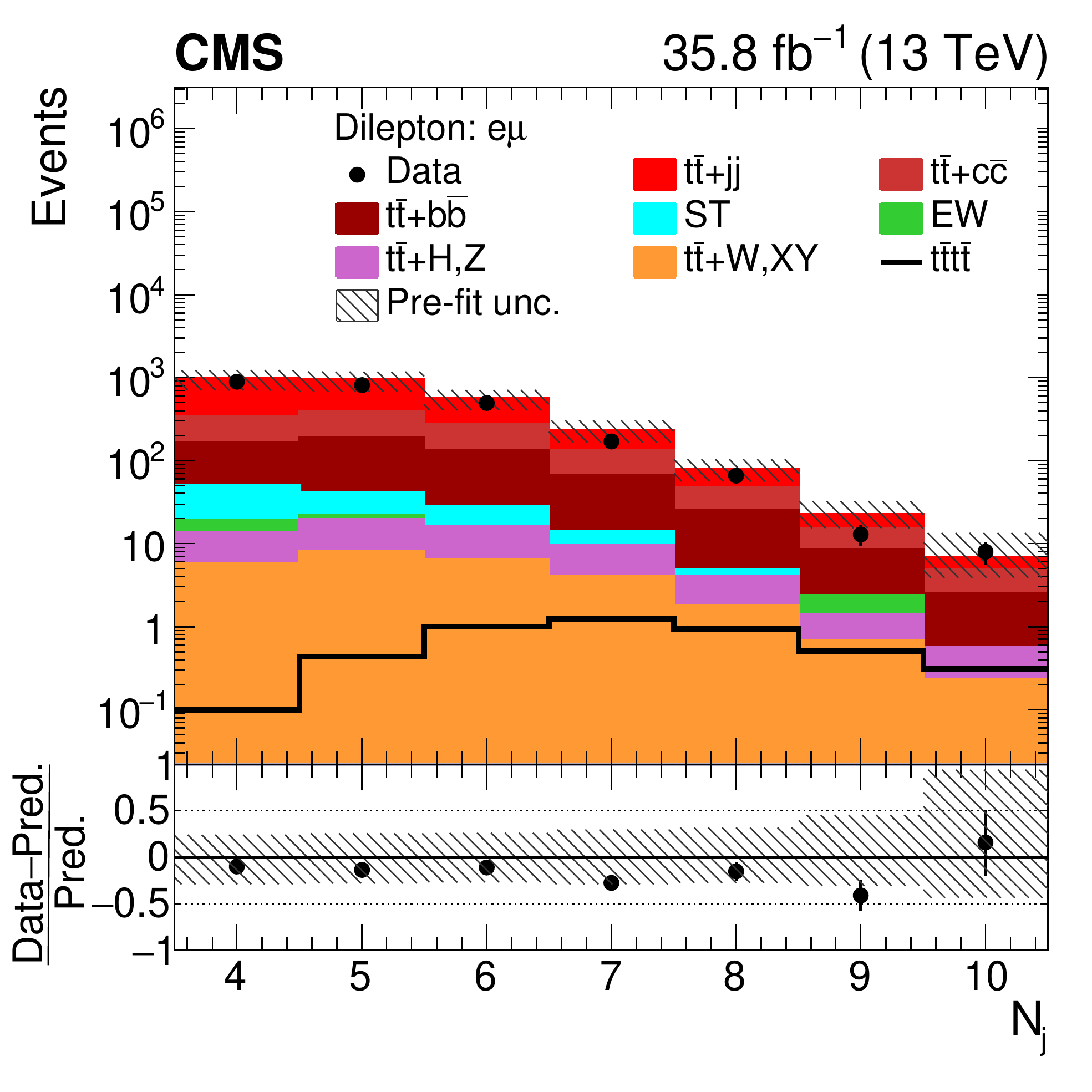}
\includegraphics[width=0.45\textwidth]{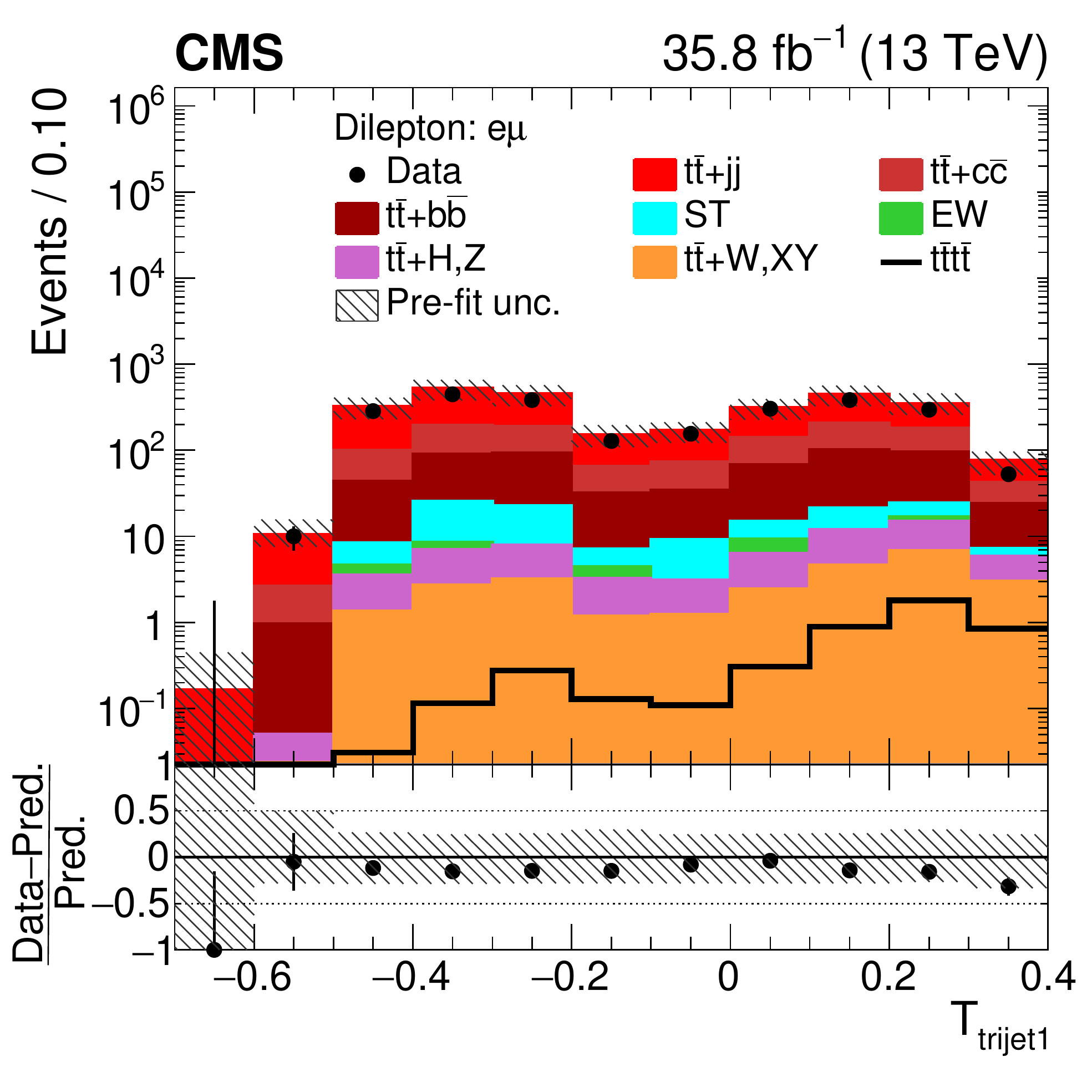}
\caption{Distributions of \njets and \BDTtrijetone in the $\MM$ (upper row) and $\PGmpm \Pemp$ (lower row) channels. In the upper panels of each figure, the data are shown as dots with error bars representing statistical uncertainties, MC simulations are shown as a histogram. The lower panels show the relative difference between the data and the sum of all of the standard model backgrounds. In each panel, the shaded band represents the total uncertainty in the dominant \ttbar background estimate. See Section~\ref{subsec:mva} for the definitions of the variables. }
\label{fig:controldistributions_OS1}
\end{figure}

\begin{figure}[ht!]
\centering
\includegraphics[width=0.45\textwidth]{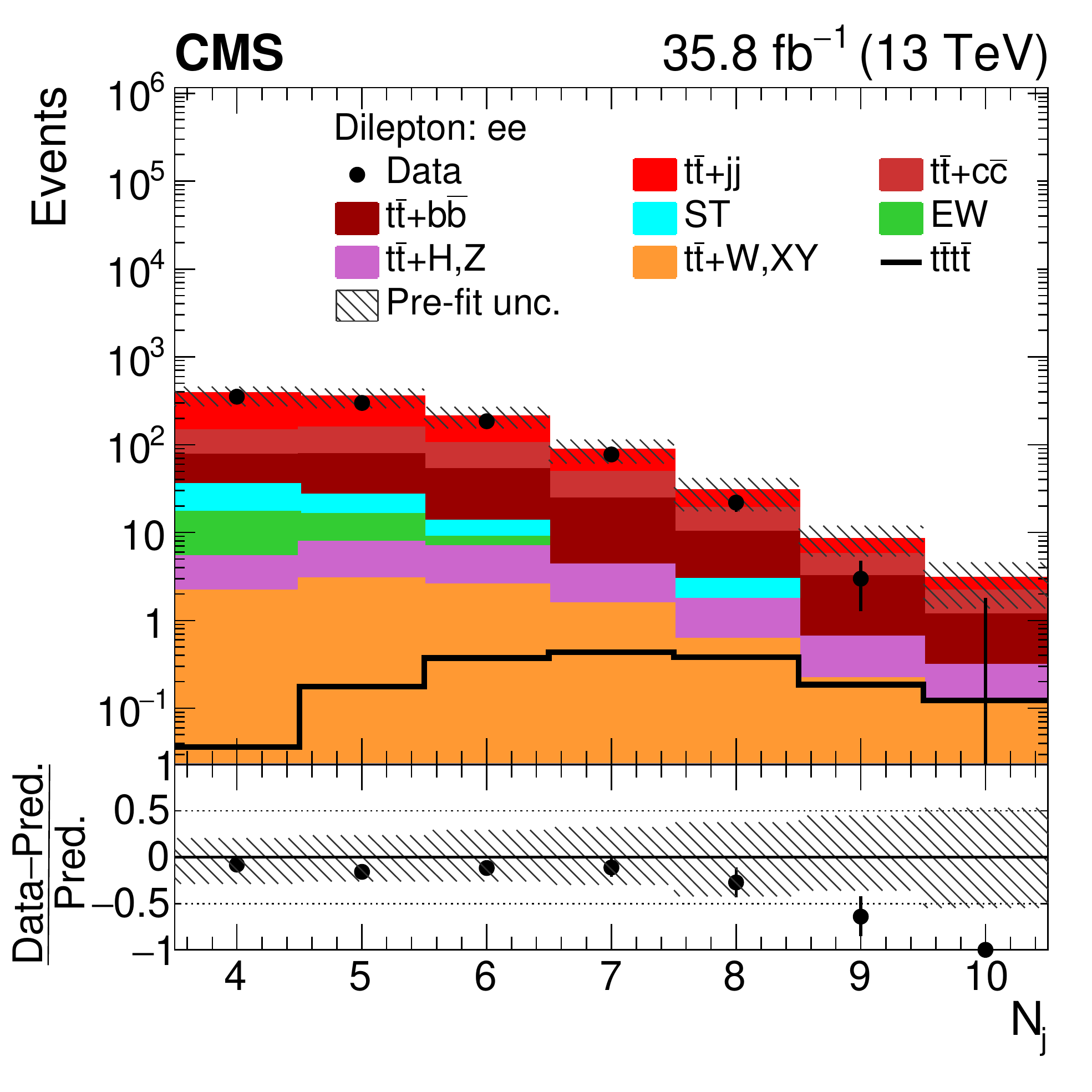}
\includegraphics[width=0.45\textwidth]{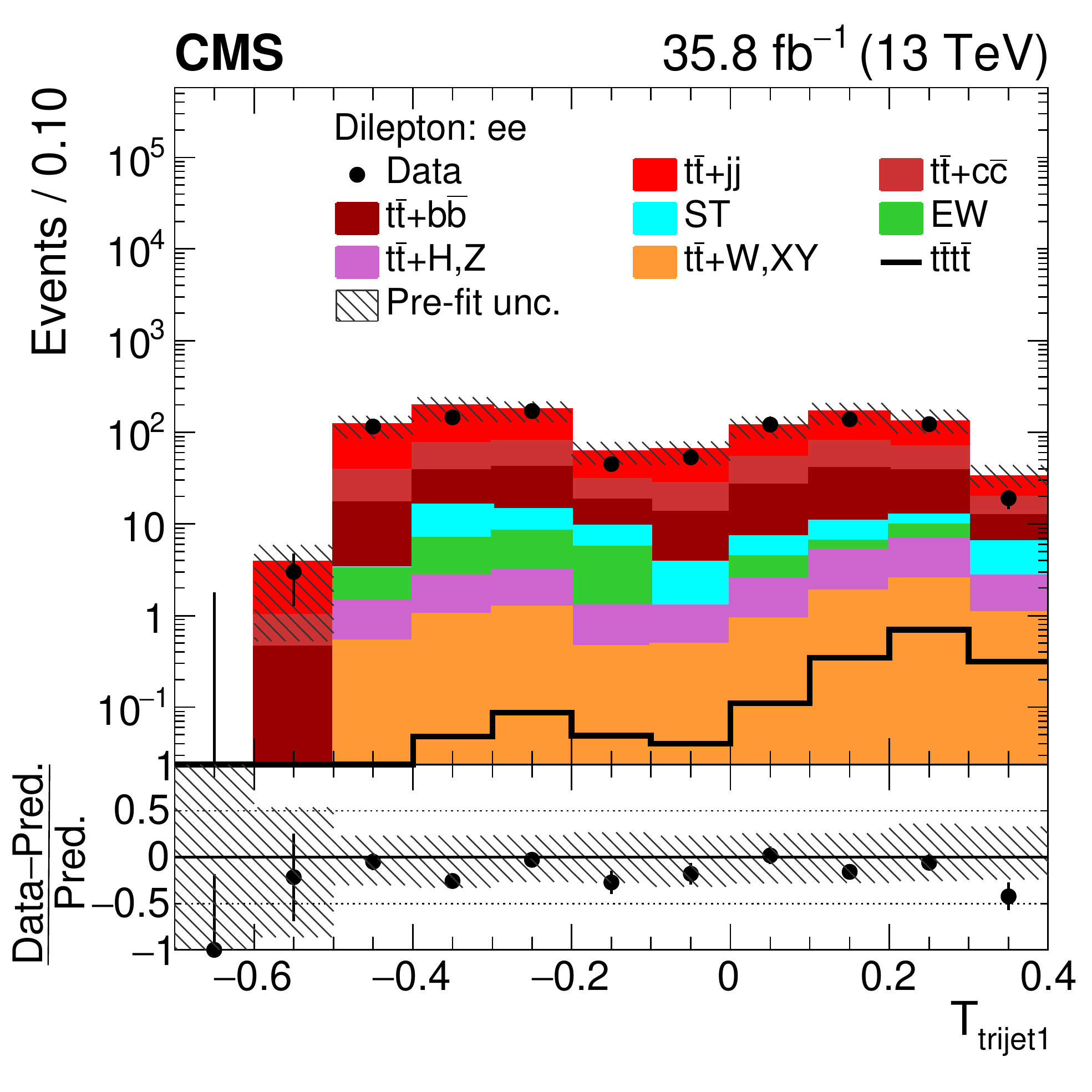}
\caption{Distributions of \njets and \BDTtrijetone in the $\EE$ channel.
In the upper panels of each figure, the data are shown as dots with error bars representing statistical uncertainties, MC simulations are shown as a histogram. The lower panels show the relative difference between the data and the sum of all of the standard model backgrounds. In each panel, the shaded band represents the total uncertainty in the dominant \ttbar background estimate. See Section~\ref{subsec:mva} for the definitions of the variables. }
\label{fig:controldistributions_OS2}
\end{figure}

Figures~\ref{fig:controldistributions_SL1}--\ref{fig:controldistributions_OS2}
show the comparison of the data and simulations after these selections
have been applied for both the single-lepton and dilepton analyses.
The simulation of \ttbarjets process is split into three categories:
top quark pair associated with two additional light flavor or gluon
jets ({\ttbar}+jj), top quark pair associated with a charm quark pair
({\ttbar}+{$\ccbar$}), and top quark pair associated with a bottom quark
pair ({\ttbar}+{$\bbbar$})~\cite{Sirunyan:2018mvw}.
The definitions of the variables in the figures are given in the next section.

\subsection{Multivariate discriminants}
\label{subsec:mva}
Boosted decision trees (BDTs)~\cite{bdt1,bdt2} are used in two roles in this analysis: to identify the top quarks and to improve the discrimination between signal and background.
The jet multiplicity, jet properties and the number of the {\cPqb}
jets, as well as associated kinematic variables, feature strongly in
the choice of BDT input variables. The method is based on the
strategies developed for the previous 13\TeV CMS analyses in the
single-lepton and opposite-sign dilepton final
states~\cite{CMStttt2015}.
All BDTs are trained using the \textsc{AdaBoost} algorithm~\cite{Freund:1997xna}, as implemented in the \textsc{tmva} package~\cite{tmva}, and return a discriminant as output.

The BDT for identifying hadronically decaying top quarks classifies
combinations of three jets (trijet) on how consistent they are with
the trijet originating from the all-hadronic decay of a top quark,
rather than from other sources such as initial-state radiation (ISR)
or final-state radiation (FSR).
Its input variables consist of the invariant dijet and trijet masses,
the {\cPqb} tagging information for the jet not associated to the dijet, and the angles between the three
jets. This BDT is trained to distinguish between the three jets from a hadronically decaying top quark and any other permutation of 3-jet combinations using the ME information in tt+jets simulations.

Because of the high jet multiplicity in both signal and background events, many three-jet combinations are possible. The trijet permutations for each event are ranked according to their discriminant value, from highest to lowest.
In the single-lepton channel, each \ttbar background event contains a genuine hadronic top quark decay, so the jets included in the first-ranked trijet (\BDTtrijetone) are removed and the highest-ranked discriminant using the remaining jets (\BDTtrijettwo) is used.
In the dilepton channels, the \ttbar background contains no hadronic top quark decays, so only the output for \BDTtrijetone is used as the discriminant.

The BDTs, yielding the discriminants for the single-lepton channel
(\BDTljets) and for the dilepton channel (\BDTdilep), use the
discriminant from the trijet associations, described above, as one of
its input variables.
In the single-lepton channel, \BDTljets is trained separately for each
jet multiplicity, and inclusively over the number of {\cPqb}-tagged jets.
In the dilepton channel, the training is done unitarily for all jet
multiplicities while separately in $\MM, \PGmpm \Pemp$, and $\EE$
states.
 The choice of input variables is optimized separately for the two channels and is based on the characteristics of the lepton and jet activity in the events. The resulting variable lists are different for the two channels. The variables can be grouped into three categories: event activity, event topology, and {\cPqb} quark multiplicity. Although many of the input variables are correlated, each one contributes some additional discrimination between the \ttbar background and the \tttt signal.

Studies of the differences between the simulated \ttbar and \tttt events have led to the selection of the following variables describing the hadronic activity in the event:
\begin{enumerate}

\item The number of jets present in the event, \njets.

\item The scalar sum of the \pt of all medium working point {\cPqb} jets in the event, \htb.

\item The ratio of the sum \HT of the four highest \pt jets in the event in the single-lepton channel, or the two jets with the highest {\cPqb} tagging discriminant in the dilepton channel, to the \HT of the other jets in the event, \htrat.

\item The \HT sum in the event, subtracting the scalar \pt sum of the two highest {\pt} {\cPqb} jets, \httwom.

\item The transverse momenta of the jets with the third- and fourth-largest \pt in the event, \thirdjetpt and \fourthjetpt.

\item The reduced event mass, \redhadmass, defined as the invariant mass of the system comprising all the jets in the reduced event,
where the reduced event is constructed by removing the jets contained in \BDTtrijetone in single-lepton events.
In \ttbar events, the reduced event will typically only contain the {\cPqb} jet from the semileptonic top quark decay and jets arising from ISR and FSR. Conversely, a reduced \tttt event can contain up to two hadronically decaying top quarks and, as a result, a relatively high reduced event mass.

\item The reduced event \HT, \htx, is defined as the \HT of all jets in the single-lepton event selection excluding those contained in \BDTtrijetone .

\end{enumerate}

The event topology is characterized by the two variables:
\begin{enumerate}
\item Event sphericity, $S$,~\cite{sphericity}, calculated from all of the jets in the event in terms of the normalized tensor
\begin{linenomath}
$M^{\alpha \beta} = {\sum_i p_i^\alpha p_i^\beta}/{\sum_i \abs{\vec{p_i}}^2} \label{eq:sphTensor},$
\end{linenomath}
where $\alpha$ and $\beta$ refer to the three-components of the momentum of the $i$th jet. The sphericity is defined as $S = (3/2)(\lambda_2+\lambda_3)$, where $\lambda_2$ and $\lambda_3$ are the two smallest eigenvalues of $M^{\alpha \beta}$.
The sphericity in \tttt events should differ from that in background \ttbar events of the same energy, since the jets in \ttbar events will be less isotropically distributed because of their recoil from sources such as ISR.

\item Hadronic centrality, $C$, defined as the value of \HT divided by the sum of the energies of all jets in the event.
\end{enumerate}

Since all these variables rely only on the hadronic information in the event,
sensitivity to the lepton information is provided through the \pt and $\eta$ of the
highest \pt lepton (or the only lepton for the single-lepton channel) $\left(\leadleppt,\leadlepeta\right)$ and the angular difference (\drll) between the leptons in dilepton events.
The {\cPqb} jet multiplicity is characterized in terms of the number of {\cPqb} jets tagged by the CSVv2 algorithm operating
at its loose (\nltags) and medium (\nmtags) operating points, and the angular separation \drbb between the {\cPqb}-tagged jets with the highest CSVv2 discriminants. Finally, the third- and fourth-highest {\cPqb} tagging discriminant values are used as they allow separation between \ttbar+light jets, and genuine additional heavy-flavor jets, as present in \tttt events.

The training variables were not changed as a function of final state or jet multiplicity.
In the single-lepton channel, the optimal variable set, listed in the order of their discriminating power, is \BDTtrijettwo; \htb; \hth; \leadleppt; \redhadmass; \htx; the third- and fourth-highest CSVv2 discriminants, and the \PT of those tagged jets; the $\PT$ for the first, second, fifth, and sixth jet.
In the dilepton channel, the optimal variable set, listed in the order of their discriminating power, is \njets, \BDTtrijetone, \httwom, \fourthjetpt, \nltags, \htrat, \htb, \eventsph, \drbb, \nmtags, \drll, \hth, \thirdjetpt, \leadleppt and \leadlepeta.
The MC modeling of the individual observables utilized in the discriminants \BDTljets and \BDTdilep was verified using samples of \ttbar events and found to be in agreement with the data for all the jet and {\cPqb} jet multiplicities.

 \section{Systematic uncertainties}
\label{sec:systematics}
\label{systematics}
The systematic uncertainties that affect this analysis can change the shape, or the normalization, or both, of the \BDTljets and \BDTdilep discriminants. The uncertainties are characterized in Table~\ref{tab:syst_type}. Each of the systematic uncertainty sources is modeled by one nuisance parameter. The normalization-dependent terms account for the uncertainties in the background yields, while the effect of the shape-dependent terms is evaluated using discriminant distributions whose shape has been modified by each of the uncertainties.

\begin{table}[ht!]
\begin{center}
 \topcaption{Uncertainties that affect the normalization of the data sets and shapes of the \BDTljets and \BDTdilep discriminants. Their contribution to different effects are marked by X.}
  \label{tab:syst_type}
  \medskip
\begin{tabular}{lccc}
Systematic uncertainty     & Normalization & Shape\\
\hline
Integrated luminosity         & X & \\
Pileup modeling               & X & X \\
Lepton reconstruction and identification          & X &  \\
Jet energy corrections        & X & X \\
{\cPqb} tagging               & X & X \\
Ren. and fact. scales         & X & X \\
PS scales		      & X & \\
ME-PS matching                & X &  \\
UE 		              & X &  \\
Jet multiplicity correction   & X & \\
Parton distribution functions      & X & X \\
Top quark \pt reweighting     &   & X  \\
Heavy-flavor reweighting      & X & X  \\
Rare process		      & X &  \\
\end{tabular}
\end{center}
\end{table}

The experimental uncertainties considered are:

\begin{itemize}
\item {Integrated luminosity:} A 2.5\% normalization uncertainty on the integrated luminosity~\cite{CMS:2017sdi}.

\item {Pileup modeling:} The number of pileup events in the simulation is matched to that of the data. The uncertainty due to this correction is estimated by using two sets of alternative weights derived with a variation of $\pm4.6\%$ on the total inelastic $\Pp\Pp$ cross section~\cite{ppxsec}.

\item {Lepton reconstruction and identification:} The uncertainties in lepton identification, isolation, trigger efficiencies, and tracking efficiencies were examined. After a comparison between data and simulations, we assign a normalization uncertainty of 3\% to take into account these effects.
	
\item {Jet energy corrections and resolutions:} The uncertainties due to limited knowledge of the jet energy scale (JES) and the jet energy resolution (JER) are estimated by varying the $\eta$- and $\pt$-dependent JES and JER corrections of all jets by $\pm$1 standard deviation~\cite{Khachatryan:2016kdb}. In the case of JES uncertainty, it was split into 6 components, which include uncertainties owing to the absolute jet-energy scale, the pileup offset, the extrapolation between samples of different jet-flavor composition, the parton fragmentation and underlying event modeling and residual time and $\eta$-dependent variations. Each component represents a quadratic sum of subsets of jet energy correction uncertainties from different sources. The effect of each component is evaluated separately.

\item {{\cPqb} tagging:} The uncertainty in the {\cPqb} tagging discriminant shape is estimated by varying the shape of the discriminant distribution according to its one standard deviation uncertainties in terms of the \PT, $\eta$, and flavor of the jets~\cite{Sirunyan:2018mvw}. The variations correspond to uncertainties in the jet energy scale, background contamination of the samples used to derive them, and statistical uncertainties of these data samples.

\end {itemize}
Sources of systematic uncertainties originating from theory are listed below.
\begin{itemize}

\item {Renormalization and factorization scales:} In order to estimate the uncertainty arising from missing higher-order terms in the calculation of the signal and background cross sections, renormalization and factorization scales are each modified, independently, up and down by a factor of two relative to their nominal values. The cases in which the two scales are varied in opposite directions are excluded. This is estimated for both the \ttbar and \tttt processes.

\item {Parton shower scales:} The evolution scales in the initial- and final-state PSs are separately varied by a factor of 2 and $\sqrt{2}$, respectively, up and down relative to their nominal values, in order to estimate the uncertainty attributed to the shower model. This is estimated for both the \ttbar and \tttt processes.

\item {ME-PS matching:} The uncertainty resulting from this source is estimated by varying the \POWHEG-{\textsc{box}} PS scale parameter, $h_\text{damp}$, that controls the ME and PS matching and regulates the high-$\pt$ radiation, within its uncertainty by $\pm$1 standard deviation of the measured value $h_\text{damp} = 1.581^{+0.658}_{-0.585 } m_{\PQt}$~\cite{CMS-PAS-TOP-16-021}. This is estimated for the \ttbar process.

\item {Underlying event:} The uncertainty from the UE tune of \ttbar event generator is evaluated by using simulations with varied parameters that are related to the CUETP8M2T4 tune~\cite{CUETP8M2T4_Tune}. This is estimated for the \ttbar process.

\item {Jet multiplicity correction:} The modeling of \ttbarjets production in  \POWHEG-{\textsc{box}} is insufficient to describe the data in the regions of large jet multiplicity. To allow for this, scale factors are determined from fits to the single-lepton data in the signal depleted regions ($\njets = 8$,9, and $\nmtags = 2$,3), and propagated to the signal sensitive regions. The scale factors determined in the single-lepton channel are also used in the dilepton channel taking into account the difference in the jet multiplicity between the two channels. The uncertainty resulting from this correction is 10\% for the \ttbar process~\cite{Sirunyan:2018ptc}.

\item {Parton distribution functions:} The PDF uncertainty~\cite{Butterworth:2015oua} in \ttbar production is estimated by evaluating the shape difference between the nominal simulation and simulations based on the NNPDF~\cite{Ball:2014uwa}, MMHT14~\cite{Harland-Lang:2014zoa}, and CT10~\cite{Dulat:2015mca} PDF sets. This is estimated for the \ttbar process.

\item {Top quark \pt reweighting:} The \ttbar simulation is corrected to match the observed spectra~\cite{Khachatryan:2016mnb,Sirunyan:2017mzl}. The uncertainty from the corrections made to the shape of the top quark \pt distribution is estimated by allowing the correction function to vary within a $\pm$1 standard deviation uncertainty. This is estimated for the \ttbar process.

\item {Heavy-flavor reweighting:} To correctly model the rate of additional heavy-flavor jets in \ttbar production, the uncertainty in the rate of \ttbar{+}\bbbar  is taken from the $\pm$1 standard deviation uncertainty in the measured value~\cite{ttbb}. This is estimated for the \ttbar process. As a cross-check, an independent uncertainty on {\ttbar}+{$\ccbar$} production was added. The resulting effect on the expected sensitivity of the search was found to be negligible.

\item {Rare processes:} Uncertainties from the cross sections of rare processes of \ttbar pair production in association with one or two massive gauge bosons and triple top quark production are taken into account by allowing them to vary within 50\% of their SM value~\cite{TOP-17-009}.

\end {itemize}

The simulated samples used to evaluate the PS, ME-PS and UE uncertainties are statistically limited, so these uncertainties are estimated conservatively by assigning the larger value between the statistical uncertainty of these simulated samples and the rate change of these simulated sample from the nominal simulation as uncertainty, independently for different jet multiplicities.

 \section{Results}
\label{sec:results}
A simultaneous binned maximum-likelihood template fit to the single-lepton, dilepton, and combined experimental results was used to determine the signal strength parameter, which is defined as the ratio of the observed and predicted SM \tttt cross sections, $\mu=\sigma_{\tttt}^\text{obs}/\sigma_{\tttt}^\text{SM}$. To increase the sensitivity of the analysis, events are categorized depending on their jet and {\cPqb}-tagged jet multiplicities. In the single-lepton channel these categories are: $\njets = 7$, 8, 9, and $\geq$10 and $\nmtags = 2$, 3, and $\geq$4 in each jet multiplicity region. In the dilepton channel these are $\njets = 4$--5, 6--7, and $\geq$8 and $\nmtags = 2$, and $\geq$3 in each jet multiplicity region. In each category the binning was chosen to ensure at least 4 predicted background events per bin.

The likelihood function incorporates each of the systematic
uncertainties in the signal and background \BDTdilep and \BDTljets
templates as nuisance parameters in the fit. The systematic uncertainties attributed to the trigger or specific to the jet or lepton reconstruction were treated as fully correlated among the different final states. The normalization uncertainties are included assuming a log-normal distribution for the nuisance parameters, while the shape uncertainties are included as Gaussian-distributed parameters.

\begin{figure}[hp!]
	\centering
	\includegraphics[width=\textwidth]{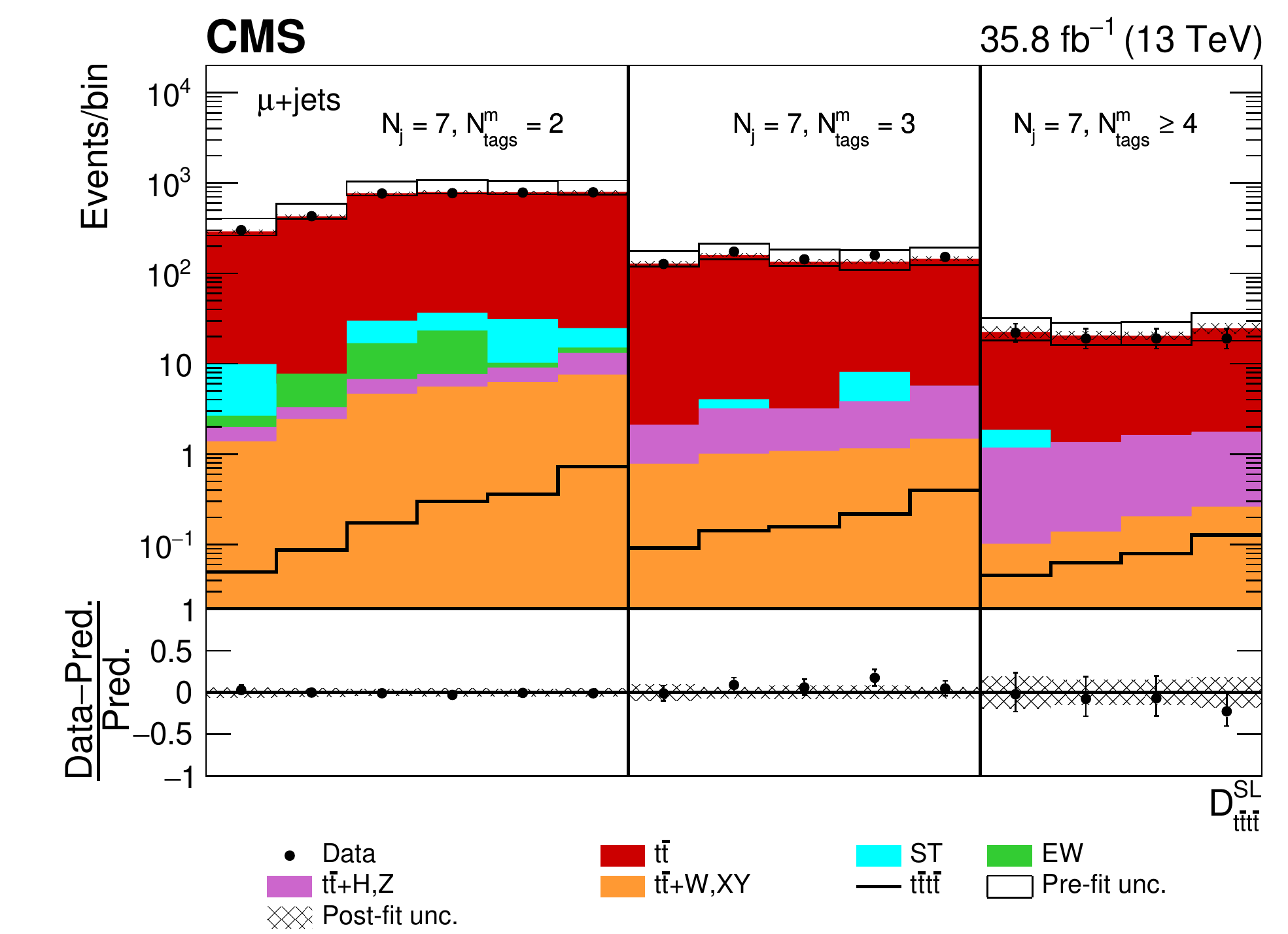}\\
	\caption{
	Post-fit \BDTljets distribution in the single-muon channel for events satisfying baseline single-lepton selection and $\njets = 7$, $\nmtags = 2$, 3, $\geq$4. Non-uniform binning of the BDT discriminant was chosen to achieve approximately uniform distribution of the \ttbar background. Dots represent data. Vertical error bars show the statistical uncertainties in data. The post-fit background predictions are shown as shaded histograms. Open boxes demonstrate the size of the pre-fit uncertainty in the total background and are centered around the pre-fit expectation value of the  prediction. The hatched area shows the size of the post-fit uncertainty in the background prediction. The signal histogram template is shown as a solid line. The lower panel shows the relative difference of the observed number of events over the post-fit background prediction. }
	\label{fig:postfit_mu7}	
\end{figure}

\begin{figure}[hp!]
	\centering
	\includegraphics[height=0.4\textheight]{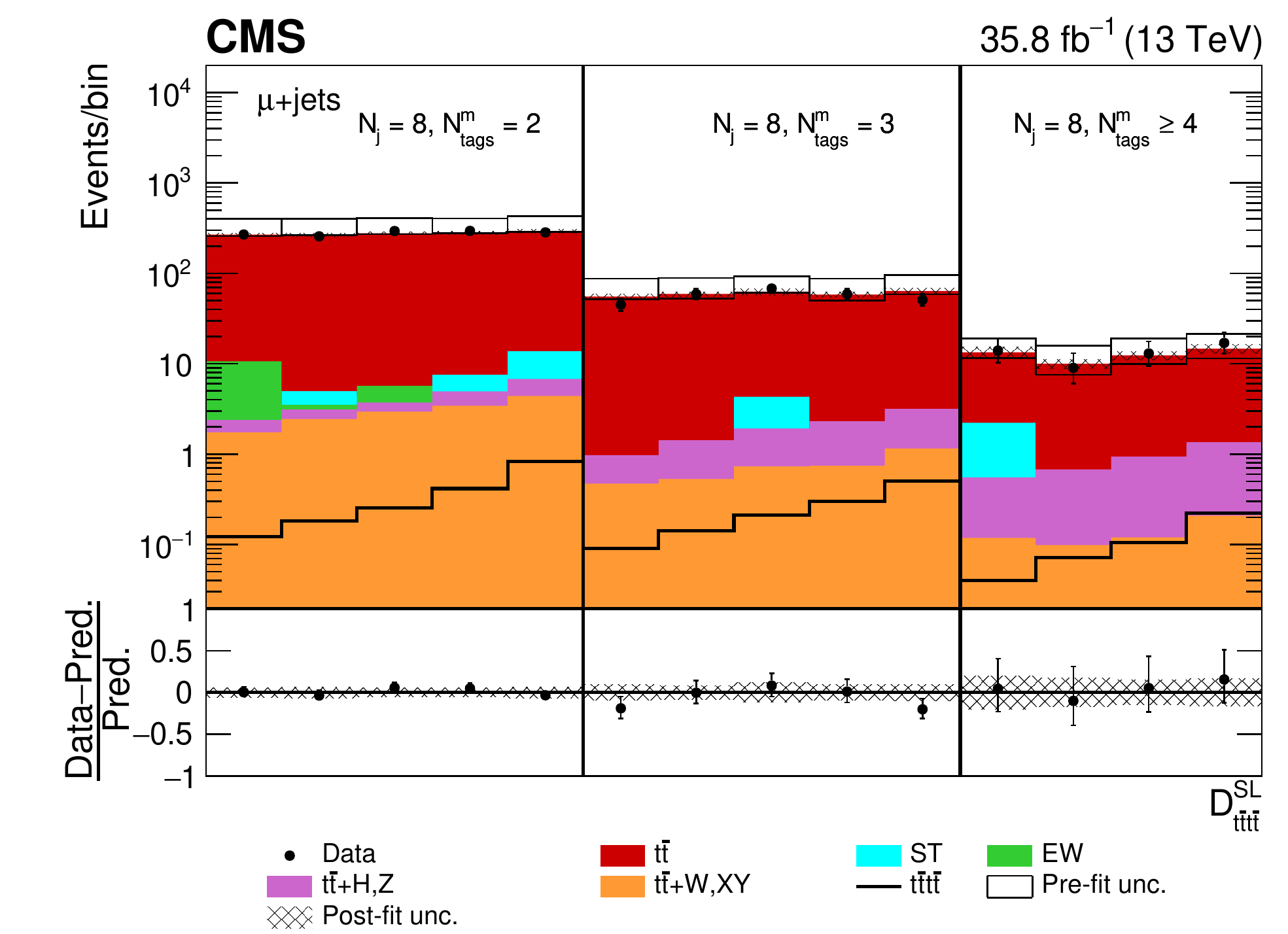}\\
	\includegraphics[height=0.4\textheight]{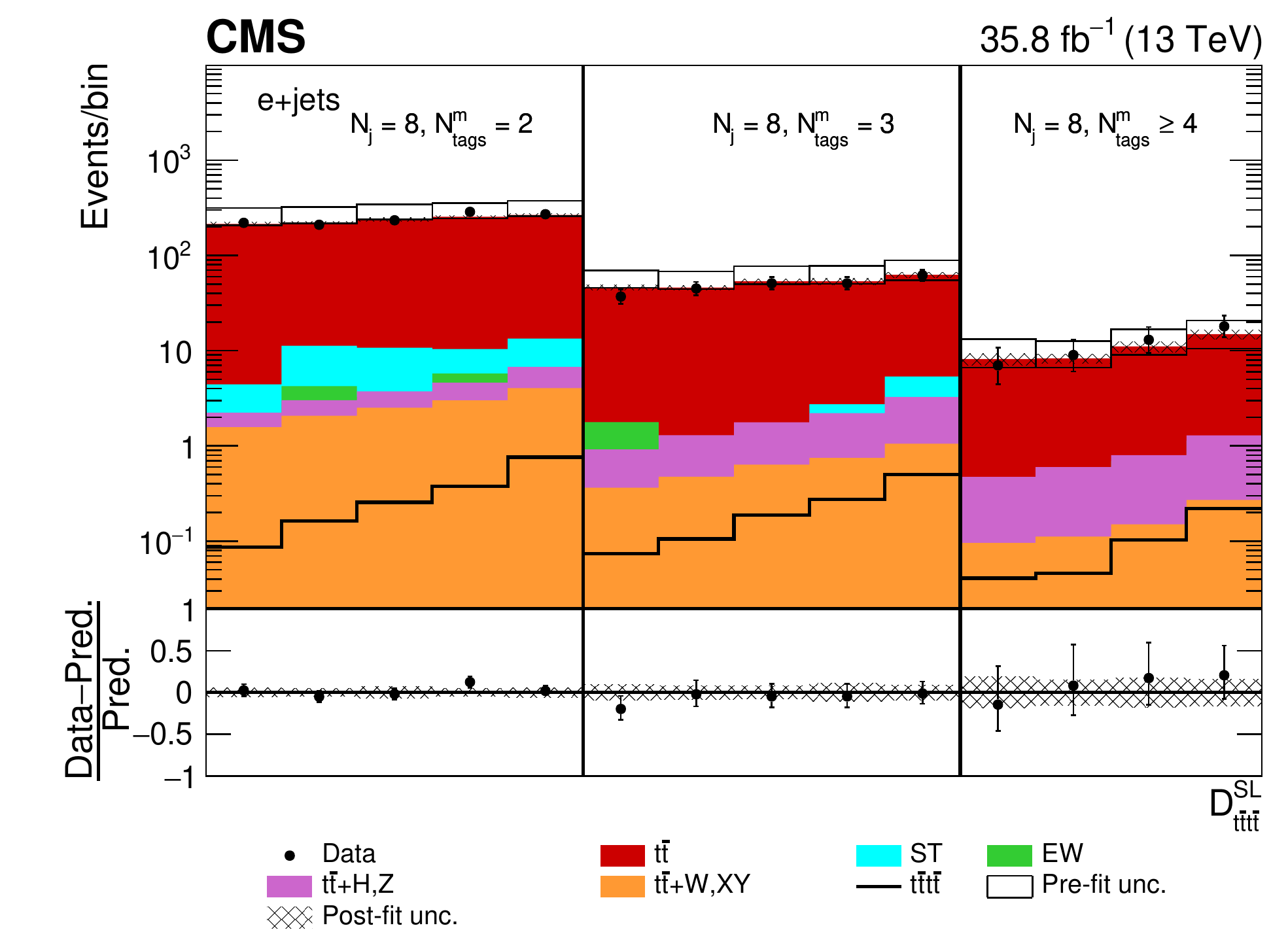}
	\caption{
	Post-fit \BDTljets distribution in the (upper row) single-muon and (lower row) single-electron channels for events satisfying baseline single-lepton selection and $\njets = 8$, $\nmtags = 2$, 3, $\geq$4.  Non-uniform binning of the BDT discriminant was chosen to achieve approximately uniform distribution of the \ttbar background. Dots represent data. Vertical error bars show the statistical uncertainties in data. The post-fit background predictions are shown as shaded histograms. Open boxes demonstrate the size of the pre-fit uncertainty in the total background and are centered around the pre-fit expectation value of the  prediction. The hatched area shows the size of the post-fit uncertainty in the background prediction. The signal histogram template is shown as a solid line. The lower panel shows the relative difference of the observed number of events over the post-fit background prediction. }
	\label{fig:postfit_muel8}	
\end{figure}

\begin{figure}[hp!]
	\centering
	\includegraphics[height=0.4\textheight]{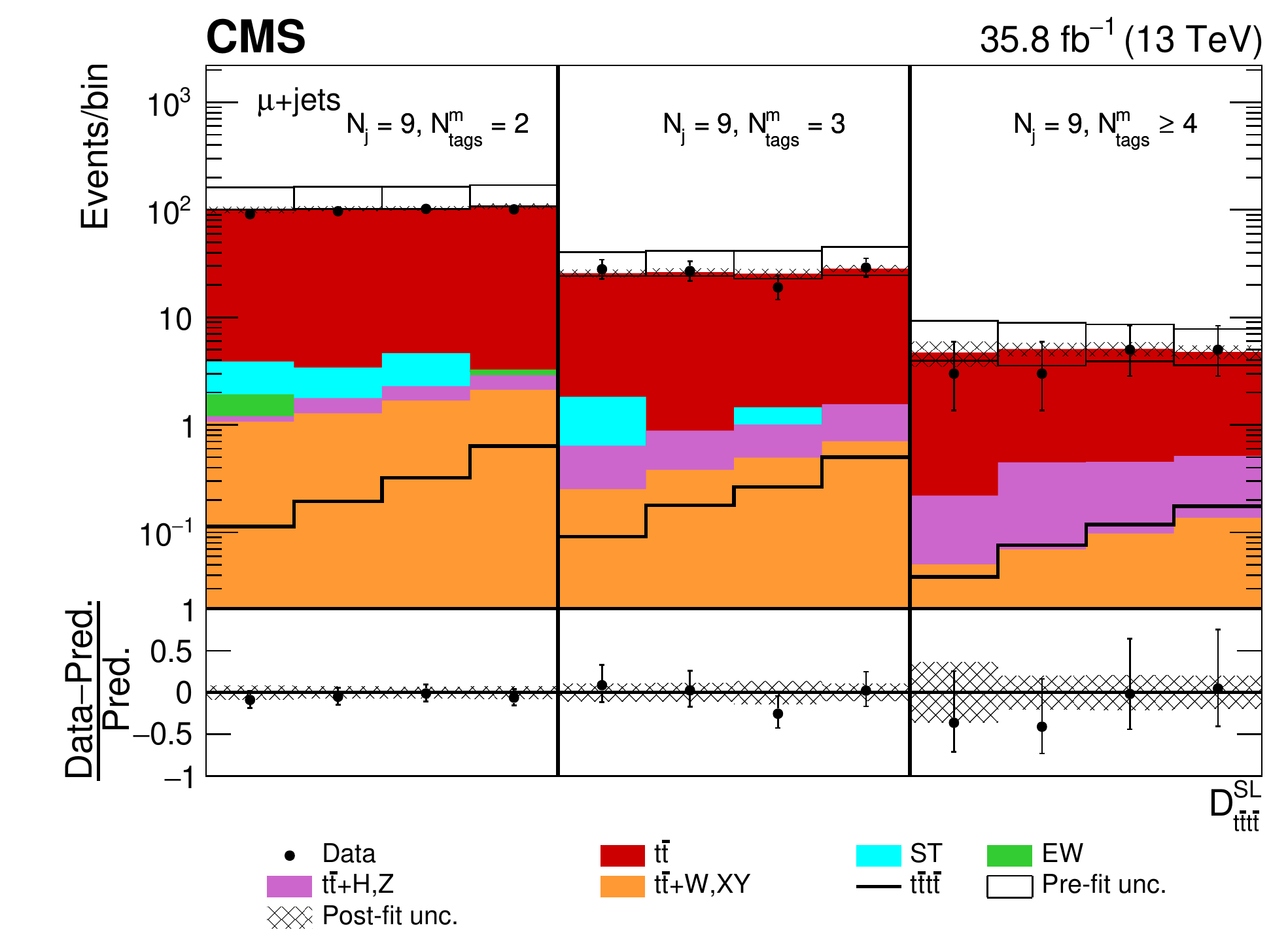}\\
	\includegraphics[height=0.4\textheight]{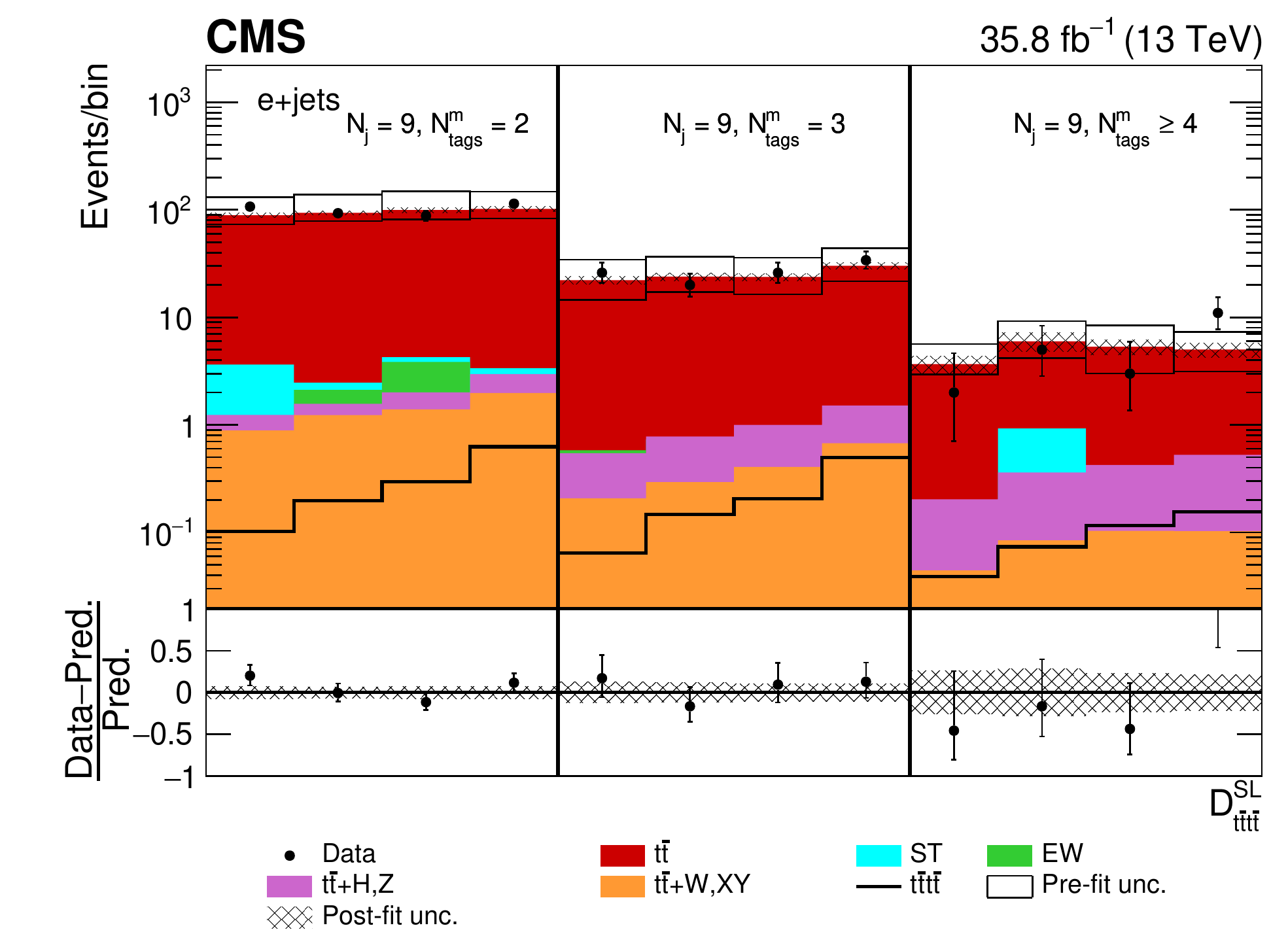}
	\caption{
	Post-fit \BDTljets distribution in the (upper row) single-muon and (lower row) single-electron channels for events satisfying baseline single-lepton selection and $\njets = 9$, $\nmtags = 2$, 3, $\geq$4.  Non-uniform binning of the BDT discriminant was chosen to achieve approximately uniform distribution of the \ttbar background. Dots represent data. Vertical error bars show the statistical uncertainties in data. The post-fit background predictions are shown as shaded histograms. Open boxes demonstrate the size of the pre-fit uncertainty in the total background and are centered around the pre-fit expectation value of the  prediction. The hatched area shows the size of the post-fit uncertainty in the background prediction. The signal histogram template is shown as a solid line. The lower panel shows the relative difference of the observed number of events over the post-fit background prediction. }
	\label{fig:postfit_muel9}	
\end{figure}

\begin{figure}[hp!]
	\centering
	\includegraphics[height=0.4\textheight]{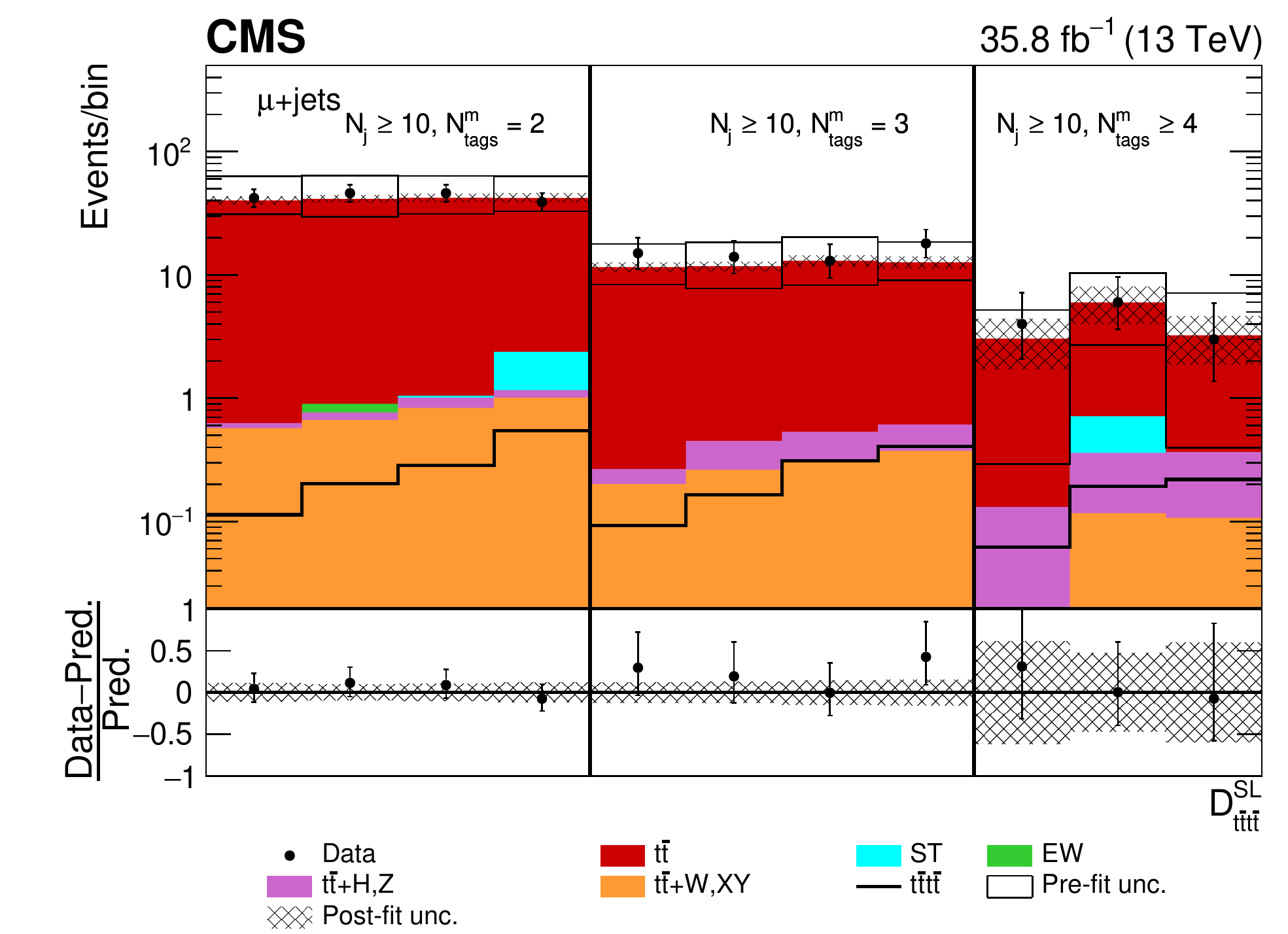}\\
	\includegraphics[height=0.4\textheight]{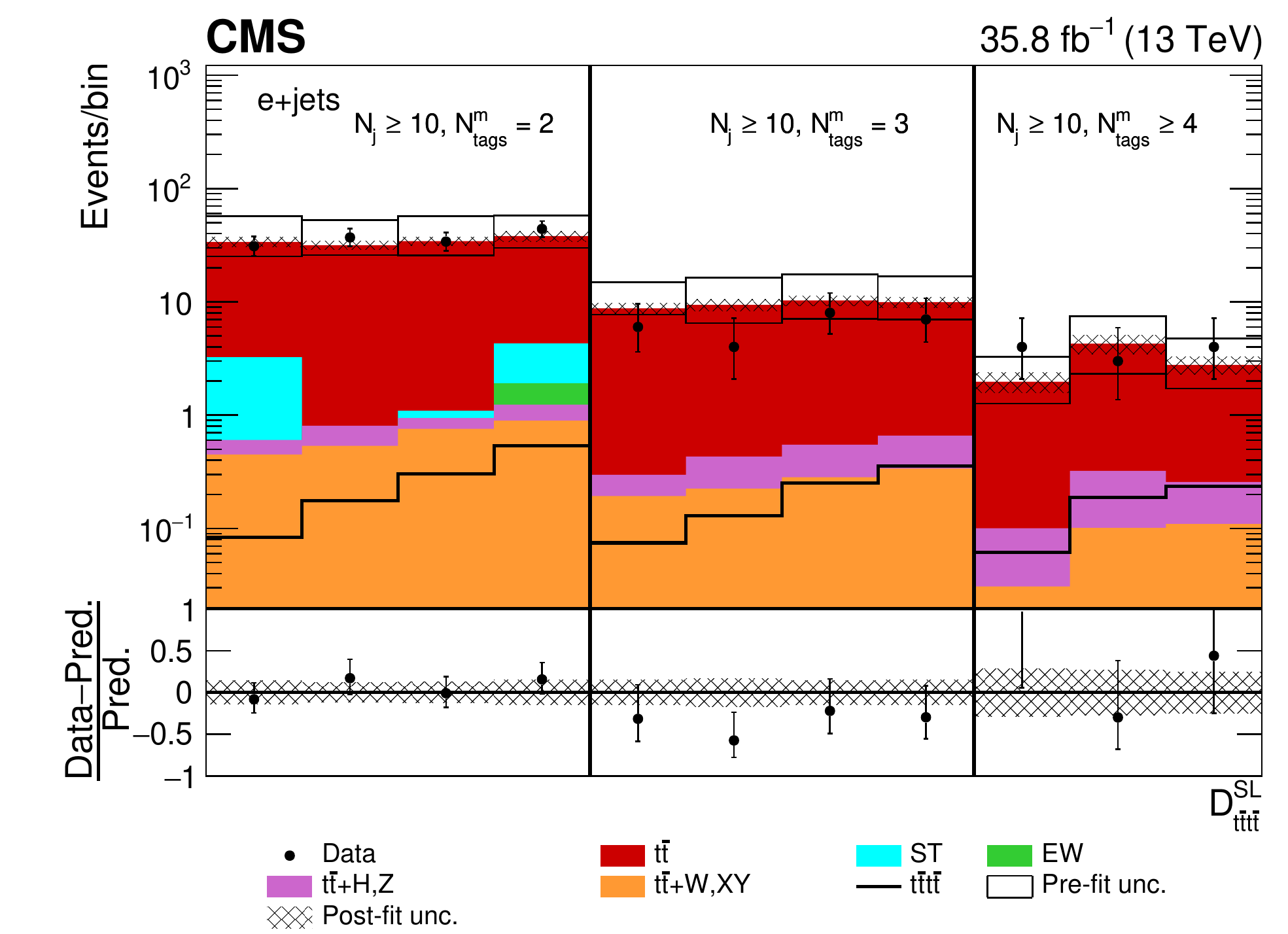}
	\caption{
    Post-fit \BDTljets distribution in the (upper row) single-muon and (lower row) single-electron channels for events satisfying baseline single-lepton selection and $\njets \geq 10$, $\nmtags = 2$, 3, $\geq$4.  Non-uniform binning of the BDT discriminant was chosen to achieve approximately uniform distribution of the \ttbar background. Dots represent data. Vertical error bars show the statistical uncertainties in data. The post-fit background predictions are shown as shaded histograms. Open boxes demonstrate the size of the pre-fit uncertainty in the total background and are centered around the pre-fit expectation value of the  prediction. The hatched area shows the size of the post-fit uncertainty in the background prediction. The signal histogram template is shown as a solid line. The lower panel shows the relative difference of the observed number of events over the post-fit background prediction. }
	\label{fig:postfit_muel10}	
\end{figure}

\begin{figure}[hp!]
        \centering
        \includegraphics[height=0.4\textheight]{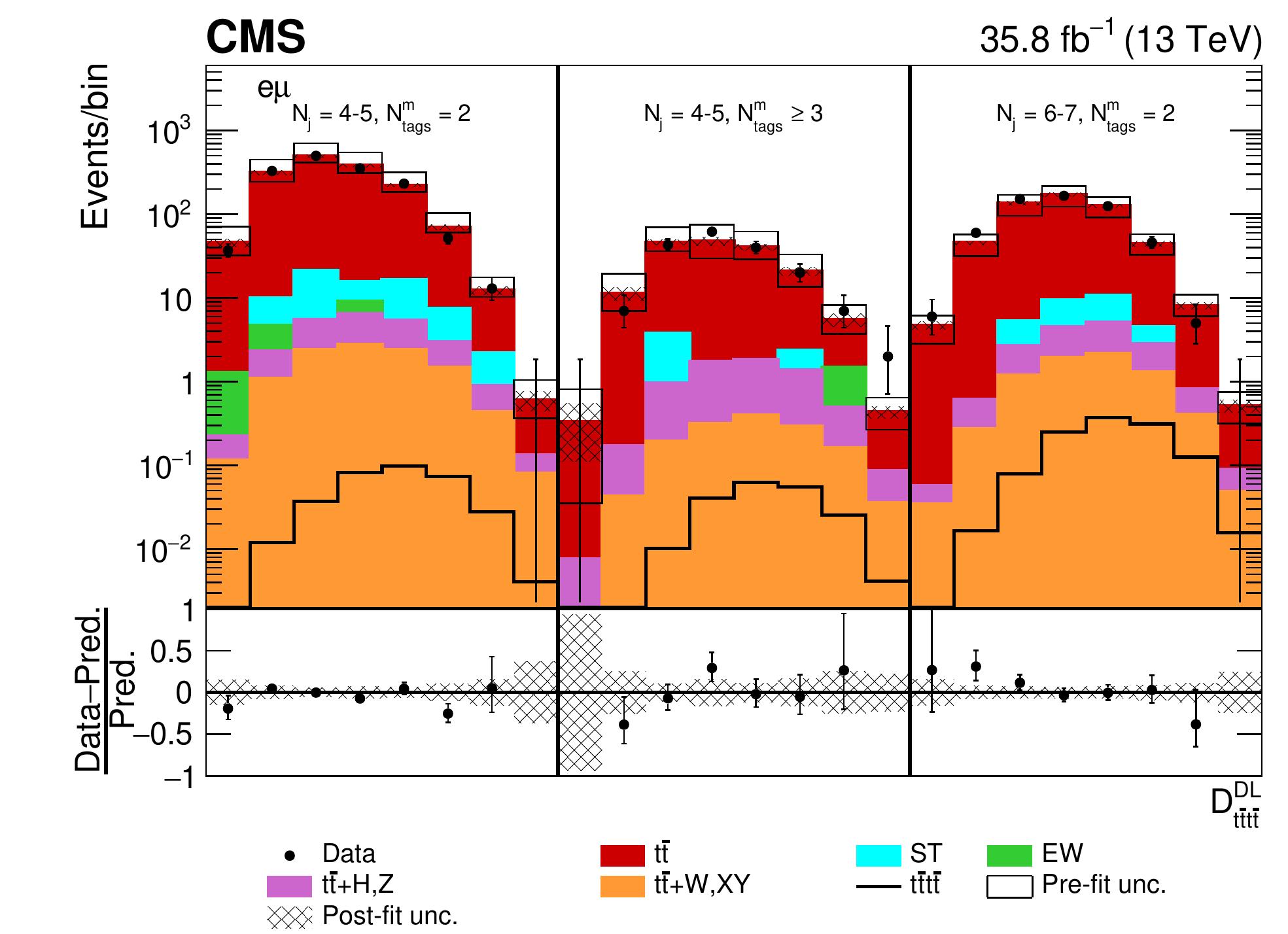}\\
        \includegraphics[height=0.4\textheight]{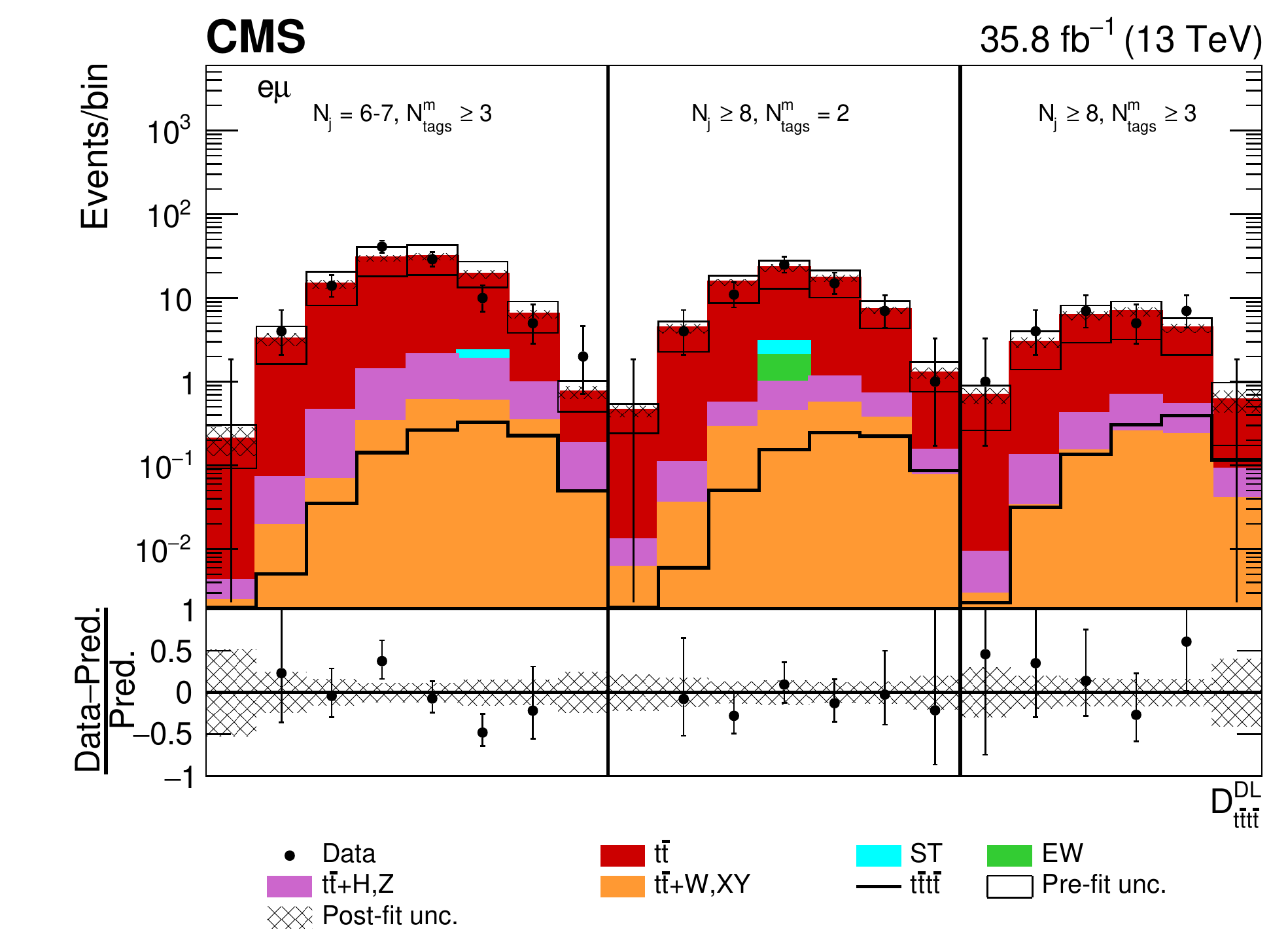}
        \caption{
        \label{fig:postfit_dileptonsmuel}
        Post-fit \BDTdilep distributions in the $\PGmpm \Pemp$ channel for events satisfying baseline opposite-sign dilepton selection and (upper row) $\njets = 4$--5, $\nmtags = 2$, $\geq$3, $\njets = 6$--7, $\nmtags = 2$ and (lower row) $\njets = 6$--7, $\nmtags \geq 3$, $\njets \geq 8$, $\nmtags = 2$, $\geq$3. Dots represent data. Vertical error bars show the statistical uncertainties in data. The post-fit background predictions are shown as shaded histograms. Open boxes demonstrate the size of the pre-fit uncertainty in the total background and are centered around the pre-fit expectation value of the  prediction. The hatched area shows the size of the post-fit uncertainty in the background prediction. The signal histogram template is shown as a solid line. The lower panel shows the relative difference of the observed number of events over the post-fit background prediction. }
\end{figure}

\begin{figure}[hp!]
        \centering
        \includegraphics[height=0.4\textheight]{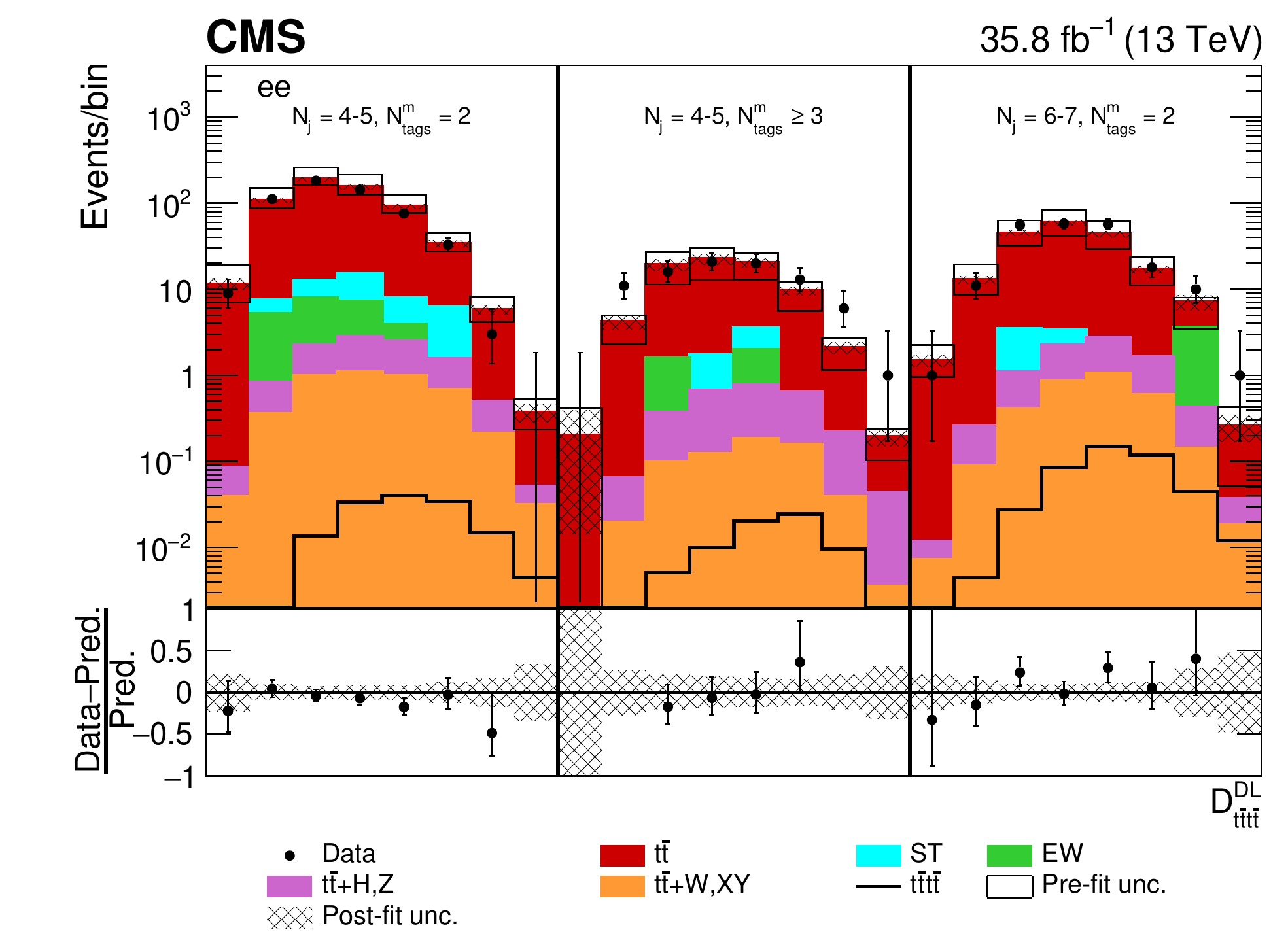}\\
        \includegraphics[height=0.4\textheight]{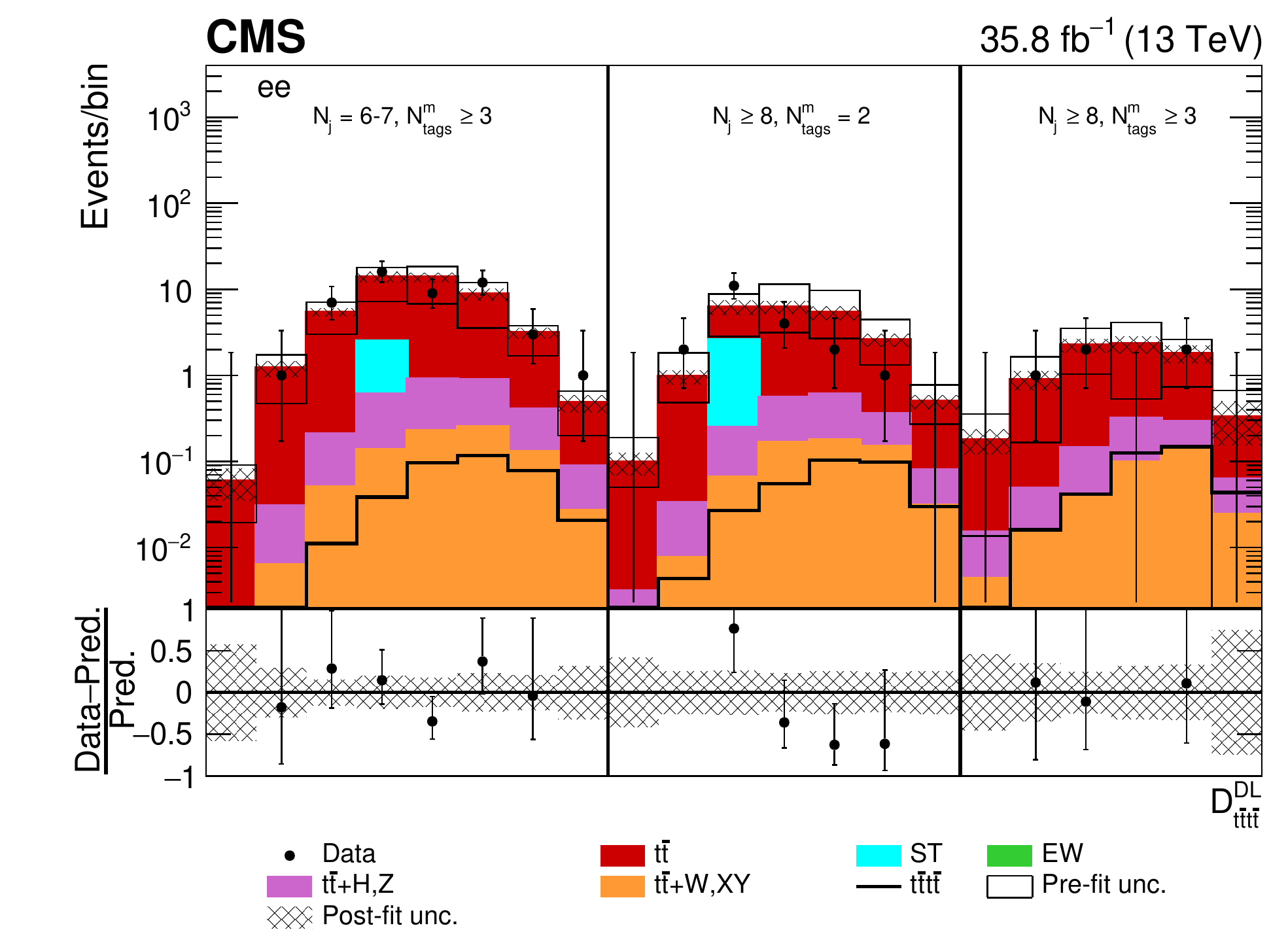}
        \caption{
        \label{fig:postfit_dileptonselel}
        Post-fit \BDTdilep distributions in the $\EE$ channel for events satisfying baseline opposite-sign dilepton selection and (upper row) $\njets = 4$--5, $\nmtags = 2$, $\geq$3, $\njets = 6$--7, $\nmtags = 2$ and (lower row) $\njets = 6$--7, $\nmtags \geq 3$, $\njets \geq 8$, $\nmtags = 2$, $\geq$3. Dots represent data. Vertical error bars show the statistical uncertainties in data. The post-fit background predictions are shown as shaded histograms. Open boxes demonstrate the size of the pre-fit uncertainty in the total background and are centered around the pre-fit expectation value of the  prediction. The hatched area shows the size of the post-fit uncertainty in the background prediction. The signal histogram template is shown as a solid line. The lower panel shows the relative difference of the observed number of events over the post-fit background prediction. }
\end{figure}

All of the post-fit nuisance parameter values were found to be consistent with their initial values to well within their quoted uncertainties, indicating the consistency of the fit model with the observed data. Two of the post-fit nuisance parameters are significantly constrained by the fit. These correspond to the heavy-flavor reweighting and initial-state parton-shower radiation scale, which are reduced by 65\% and 30\%, respectively.
The sensitivity of the analysis is affected almost equally by the
statistical uncertainty and the combined systematic uncertainties. The
leading sources of systematic uncertainty are the {\ttbar}+heavy-flavor production reweighting, the jet multiplicity
correction, and the PS and UE modeling in \ttbar simulation.
The post-fit distributions in signal-enriched \njets and \nmtags categories are shown in Figs.~\ref{fig:postfit_mu7}--\ref{fig:postfit_muel10} for the single-lepton channel and Figs.~\ref{fig:postfit_dileptonsmuel}--\ref{fig:postfit_dileptonselel} for the dilepton channel.

\begin{figure}[hp!]
        \centering
        \includegraphics[height=0.4\textheight]{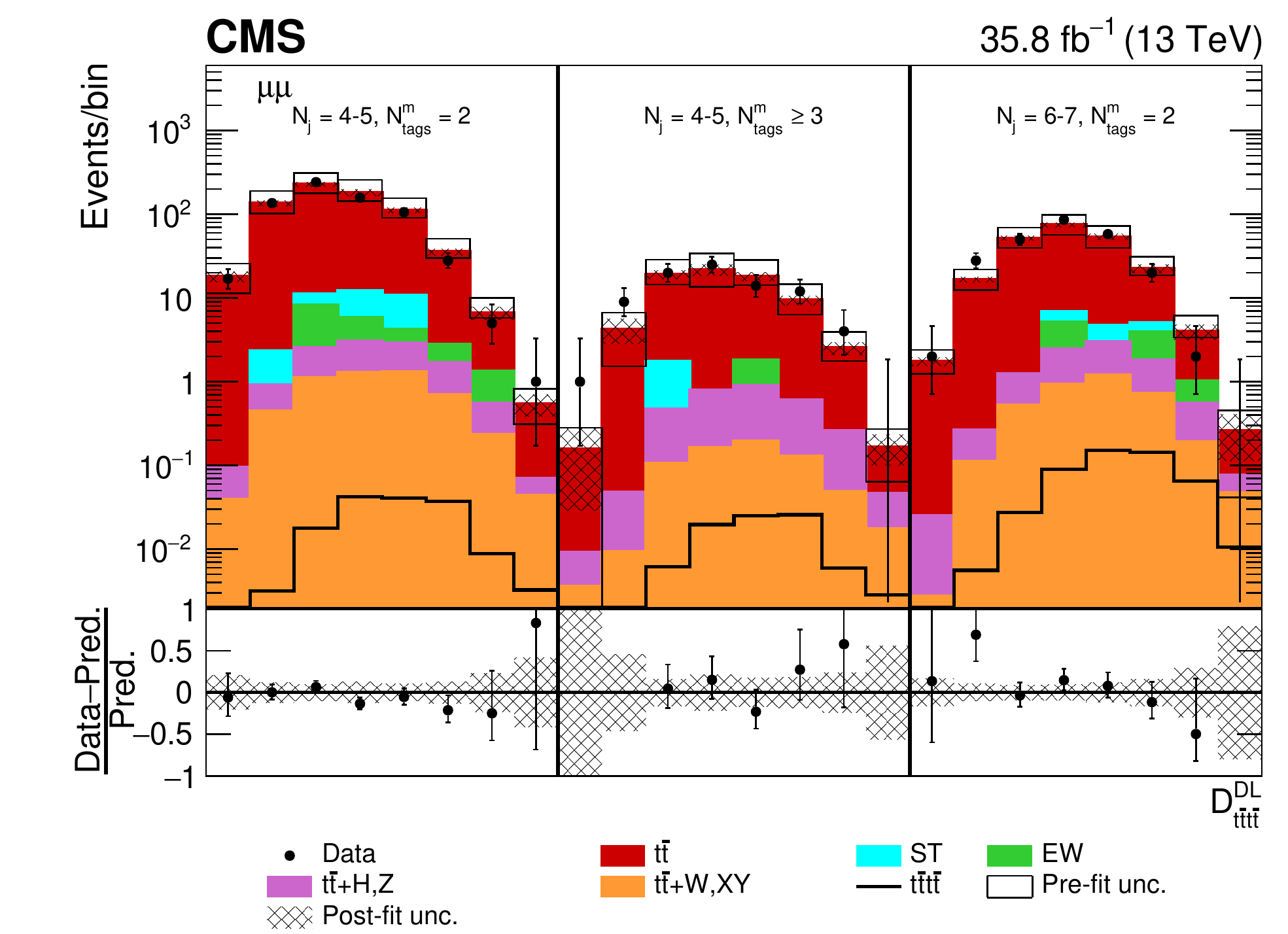} \\
        \includegraphics[height=0.4\textheight]{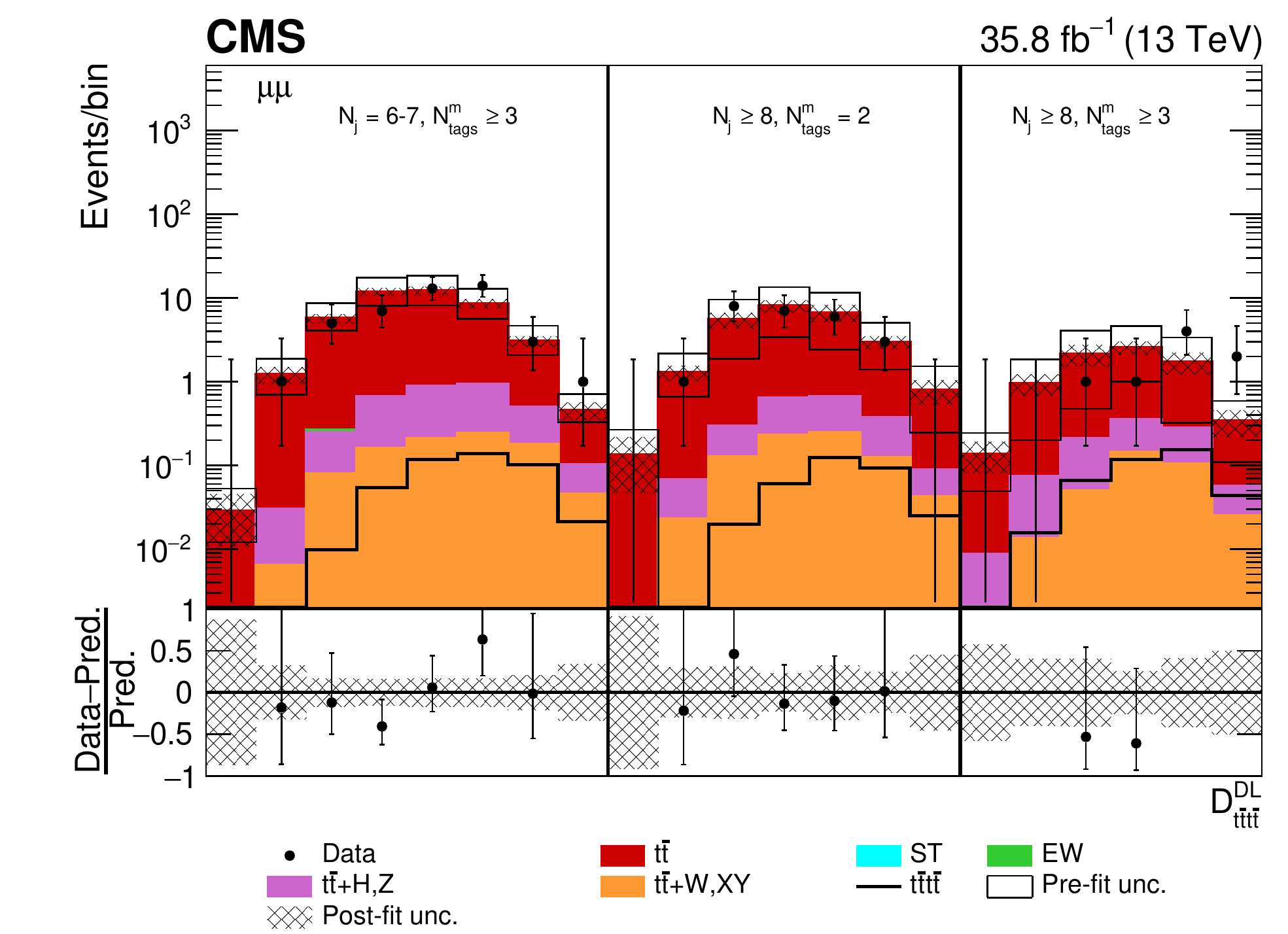}
        \caption{
        \label{fig:postfit_dileptonsmumu}
        Post-fit \BDTdilep distributions in the $\MM$ channel for events satisfying baseline opposite-sign dilepton selection and (upper row) $\njets = 4$--5, $\nmtags = 2$, $\geq$3, $\njets = 6$--7, $\nmtags = 2$ and (lower row) $\njets = 6$--7, $\nmtags \geq 3$, $\njets \geq 8$, $\nmtags = 2$, $\geq$3. Dots represent data. Vertical error bars show the statistical uncertainties in data. The post-fit background predictions are shown as shaded histograms. Open boxes demonstrate the size of the pre-fit uncertainty in the total background and are centered around the pre-fit expectation value of the  prediction. The hatched area shows the size of the post-fit uncertainty in the background prediction. The signal histogram template is shown as a solid line. The lower panel shows the relative difference of the observed number of events over the post-fit background prediction. }
\end{figure}

\begin{table}[ht!]
	\topcaption{Maximum-likelihood signal strength, $\mu$, and cross
	section estimates, as well as the expected and observed
	significance of SM \tttt production. Both $\mu$ and $\sigma_{\tttt}$ are constrained to be positive. The results for the two
	analyses from this paper are shown separately and
	combined. The results from a previous CMS multilepton measurement are also given~\cite{TOP-17-009}. The values quoted for the uncertainties on the signal strengths and cross sections are the one standard deviation (s.d.) values and include all statistical and systematic uncertainties. The expected significance is calculated assuming that the data are distributed according to the prediction of the SM with nominal \tttt production cross section value $\sigma_{\tttt}^\text{SM}$, which corresponds to the assumed signal strength modifier value $\mu=1$.}	
	\label{tab:bestfit}
	\centering
	\cmsTable{
        \begin{tabular}{ l  c   c  c  c }
		Channel  & Best fit $\mu$ & Best fit $\sigmatttt$ & Exp. significance  & Obs. significance \T \\
			 &  		& (fb) & s.d. & s.d. \T \B \\ \hline
		Single-lepton  & $\bestmusinglepton^{\,+\,\bestmusingleptonup}_{\,-\,\bestmusingleptondown}$ & $\xsfbsinglepton^{\,+\,\xsfbsingleptonup}_{\,-\,\xsfbsingleptondown}$ & $\signiffbsingleptonexp$ & $\signiffbsinglepton$   \T \B  \\
		OS dilepton  & $\bestmudilepton^{\,+\,\bestmudileptonup}$ & $\xsfbdilepton^{\,+\,\xsfbdileptonup}$ & $\signiffbdileptonexp$ & $\signiffbdilepton$ \T   \\
		Combined  & $\bestmucombo^{\,+\,\bestmucomboup}$ & $\xsfbcombo^{\,+\,\xsfbcomboup}$  & $\signiffbcomboexp$ & $\signiffbcombo$   \T   \\
		(this analysis) & & & & \\
		[\cmsTabSkip]
		SS dilepton + multilepton  & $\bestmuss^{\,+\,\bestmussup}_{\,-\,\bestmussdown}$  & $\xsfbss^{\,+\,\xsfbssup}_{\,-\,\xsfbssdown}$ & $\signiffbssexp$ & $\signiffbss$   \T \\
		Combined & $\bestmusscombounconstrained^{\,+\,\bestmusscomboupunconstrained}_{\,-\,\bestmusscombodownunconstrained}$  & $\xsfbsscombounconstrained^{\,+\,\xsfbsscomboupunconstrained}_{\,-\,\xsfbsscombodownunconstrained}$ & $\signiffbsscomboexp$ & $\signiffbsscombo$   \T \\
		(this analysis + \cite{TOP-17-009}) &  &  &  & \\
	\end{tabular}
	}

 \end{table}

No statistically significant deviation from the SM background prediction is observed in the \BDTdilep or \BDTljets distributions. The corresponding observed and expected significance of the signal and the best fit value of the signal strength parameter are given together with the $\tttt$ cross section in Table~\ref{tab:bestfit}. In order to quantify the experimental sensitivity of the search, the median expected significance is calculated assuming that the data are distributed according to the SM prediction with a nominal \tttt production cross section value $\sigma_{\tttt}^\text{SM}$, corresponding to the signal strength modifier value $\mu=1$. An upper limit on the \tttt production cross section is derived using the asymptotic approximation of the $\CLs$ method~\cite{Junk:1999kv,Read:2002hq,RooStats,Cowan:2011js,CMS-NOTE-2011-005}. The observed and expected 95\% confidence level (\CL) upper limits from the two analyses and their combination are listed in Table~\ref{tab:limits}. The expected upper limit on the \tttt production is calculated under assumption of a background-only hypothesis, corresponding to the signal strength modifier $\mu=0$.

\begin{table*}[ht!]
	\topcaption{Expected and observed 95\% \CL upper limits on SM
	\tttt production as a multiple of \sigmattttSM and
	in~fb.  The results for the two analyses from this
	paper are shown separately and combined. The results from a
	previous CMS multilepton search are also given~\cite{TOP-17-009}. The values quoted for the uncertainties in the expected limits indicate the regions containing 68\% of the distribution of limits expected under the background-only hypothesis. The expected upper limits are calculated assuming that the data are distributed according to the prediction of the background-only model corresponding to the scenario with signal strength modifier value $\mu=0$.}\label{tab:limits}
	\centering
	\cmsTable{
	\begin{tabular}{ l  c   c  c  c }
		Channel  & Expected limit, $\mu$  & Observed limit, $\mu$ & Expected limit  & Observed limit \T \\	
		         &                        &                       & (fb) & (fb) \T \B \\ \hline

                Single-lepton  & $\xsecmusingleptonexp^{\,+\,\xsecmusingleptonup}_{\,-\,\xsecmusingleptondown}$ & $\xsecmusinglepton$ & $\xsecfbsingleptonexp^{\,+\,\xsecfbsingleptonup}_{\,-\,\xsecfbsingleptondown}$ & $\xsecfbsinglepton$   \T \B  \\
                OS dilepton  & $\xsecmudileptonexp^{\,+\,\xsecmudileptonup}_{\,-\,\xsecmudileptondown}$ & $\xsecmudilepton$ & $\xsecfbdileptonexp^{\,+\,\xsecfbdileptonup}_{\,-\,\xsecfbdileptondown}$ & $\xsecfbdilepton$ \T   \\
                Combined  & $\xsecmucomboexp^{\,+\,\xsecmucomboup}_{\,-\,\xsecmucombodown}$ & $\xsecmucombo$  & $\xsecfbcomboexp^{\,+\,\xsecfbcomboup}_{\,-\,\xsecfbcombodown}$ & $\xsecfbcombo$   \T   \\
                (this analysis) & & & & \\
				[\cmsTabSkip]
                SS dilepton + multilepton & $2.5_{\,-\,0.8}^{\,+\,1.4}$ & $4.6$ & $21_{\,-\,7}^{\,+\,11}$  & $42$ \T \\
                Combined  & $\XSecMuComboAllexp^{\,+\,\XSecMuComboAllup}_{\,-\,\XSecMuComboAlldown}$ & $\XSecMuComboAllobs$  & $\XSecFbComboAllexp^{\,+\,\XSecFbComboAllup}_{\,-\,\XSecFbComboAlldown}$ & $\XSecFbComboAllobs$   \T   \\
                (this analysis + \cite{TOP-17-009}) &  &  &  & \\
	\end{tabular}
	}
 \end{table*}

 \subsection{Combination with the same-sign dilepton and multileptons channels}

An independent search for the SM \tttt production has been performed previously in same-sign dilepton and multilepton channels~\cite{TOP-17-009}.
This search is characterized by a different background composition which, in contrast to the single-lepton and opposite-sign dilepton searches, is composed of mainly the $\ttbar\PZ$ and $\ttbar\PW$ processes.
In order to exploit the complementarity of this analysis, a combination of the results from single-lepton, opposite-sign, same-sign and multilepton channels has been performed.
The combination is based on a binned likelihood function equal to a product of likelihood terms over all search regions considered in single-lepton, opposite-sign and same-sign dilepton and multilepton channels.

Because of different origins of the dominant background processes, the main systematic uncertainties in the three analyses are independent and can be treated as uncorrelated.
Nevertheless, the stability of the combination with respect to the assumption on the correlations between common sources of systematic uncertainty was tested by repeating the fit with and without correlations between the corresponding nuisance parameters.
The resulting changes in the signal strength and expected limit were found to be less than 1\% of the corresponding total uncertainties and were therefore not included.

The combined expected and observed 95\% \CL upper limits on the \tttt production are $\XSecFbComboAllexp^{\,+\,\XSecFbComboAllup}_{\,-\,\XSecFbComboAlldown}\unit{fb}$ and $\XSecFbComboAllobs\unit{fb}$, respectively, which is about a 10\% improvement on the precision of the measurement with respect to the multilepton analysis alone.
A summary of upper limit determinations from the individual analyses and their combination is provided in Table~\ref{tab:limits}.

\subsection{Effective field theory interpretation}

New physics may manifest itself as modified interactions of SM fields, even if the associated particles are too heavy to be directly probed at the LHC.
Such interactions can be modeled by extending the SM Lagrangian with terms involving composite operators of SM fields. Assuming that these terms preserve the gauge symmetries of the SM, possible new interactions can be classified according to their scaling dimension and the SM fields content~\cite{Buchmuller:1985jz,Grzadkowski:2010es,Hartland:2019bjb}. The EFT Lagrangian reads
\begin{linenomath}
\begin{equation}
\mathcal{L}_{\text{EFT}} = \mathcal{L}_{\text{SM}}^{\left(4\right)} + \frac{1}{\Lambda}\sum\limits_{k}C_{k}^{\left(5\right)}\mathcal{O}_{k}^{\left(5\right)} + \frac{1}{\Lambda^2}\sum\limits_{k}C_{k}^{\left(6\right)}\mathcal{O}_{k}^{\left(6\right)} + \ldots,
\label{eq:eftlagrangian}
\end{equation}
\end{linenomath}
where $\mathcal{L}_{\text{SM}}^{\left(4\right)}$ is the SM Lagrangian, while $\mathcal{O}_{k}^{\left(n\right)}$ and $C_{k}^{\left(n\right)}$ denote dimension\nobreakdash-$n$ (dim-$n$) composite operators and their coupling parameters, respectively. Each term in the sum is suppressed by $\Lambda^{n-4}$, where $\Lambda$ is an energy scale that characterizes the new physics and $n$ is the scaling dimension of the corresponding operator. The energy scale, $\Lambda$, is the scale below which on-shell effects of BSM physics can be neglected and is typically related to the mass scale of the hypothetical BSM states. The EFT approach is generic and, in principle, experimental constraints obtained within the EFT framework can be recast into bounds on parameters of any ultraviolet-complete new physics model.

The production of four top quarks is a unique signature that provides information about models that predict enhanced interactions of the third generation quarks, such as four-fermion \tttt coupling. The dim-5 operators do not contribute to \tttt production because they do not couple to top quarks~\cite{Weinberg:1979sa}.
A minimal basis of composite dim-6 operators contributing in Eq.~\eqref{eq:eftlagrangian} was derived in Ref.~\cite{Grzadkowski:2010es}.
Only a small subset of these operators lead to four top quark production at LO in the EFT perturbation series.
In a restricted scenario~\cite{AguilarSaavedra:2018nen,Degrande:2010kt}, assuming that new physics couples predominantly to the left-handed doublet and right-handed up-type quark singlet of the third generation, only four operators are expected to contribute significantly to $\tttt$ production, namely,
\begin{linenomath}
\begin{equation}
\begin{aligned}
\WILOR      &= \!\left( \cPaqt_\mathrm{R} \gamma^\mu \cPqt_\mathrm{R} \right) \!\left( \cPaqt_\mathrm{R} \gamma_\mu \cPqt_\mathrm{R} \right), \\
\WILOLONE   &= \!\left( \opaQL \gamma^\mu \opQL \right) \!\left( \opaQL \gamma_\mu \opQL \right), \\
\WILOBONE   &= \!\left( \opaQL \gamma^\mu \opQL \right) \!\left( \cPaqt_\mathrm{R} \gamma_\mu \cPqt_\mathrm{R} \right), \\
	\WILOBEIGHT &= \!\left( \opaQL \gamma^\mu T^\mathrm{A} \opQL \right) \!\left( \cPaqt_\mathrm{R} \gamma_\mu T^\mathrm{A} \cPqt_\mathrm{R} \right),
\end{aligned}
\end{equation}
\end{linenomath}
where $\opQL$ and $\cPqt_\mathrm{R}$ denote the left-handed third generation quark doublet and the right-handed top quark singlet, respectively.
The 4-fermion $\ttbar\bbbar$ operators were not included because of the negligible {\cPqb} quark parton density in the proton. Leading order predictions for the $\Pp\Pp\to\tttt$ cross section can be parameterized using the equation
\begin{linenomath}
\begin{equation}
\sigma_{\tttt} = \sigma_{\tttt}^{\text{SM}} + \frac{1}{\Lambda^2}\sum_{k}{C_{k} \sigma_{k}^{\left(1\right)}} + \frac{1}{\Lambda^4}\sum_{{j} \leq {k} }{{C_{j} C_{k} \sigma_{{j},{k}}^{\left(2\right)}}},
\label{eq:eftsigma}
\end{equation}
\end{linenomath}
where the linear terms, $C_{k} \sigma_{{k}}^{\left(1\right)}$, represent the interference of the SM production with the dim-6 EFT contribution, while the quadratic terms include two components: the square of the diagrams containing one EFT operator, and the interference term for two diagrams, each with one EFT operator. Representing $C_{k}$ as a column-vector, $\vec{\textbf{C}}$, Eq.~\eqref{eq:eftsigma} can be expressed in a matrix form as
\begin{linenomath}
\begin{equation}
	\sigma_{\tttt} = \sigma_{\tttt}^{\text{SM}} + \frac{1}{\Lambda^2}\vec{\boldsymbol{C}}^{\boldsymbol{T}} \cdot \boldsymbol{\vec{\sigma}}^{\left(1\right)} + \frac{1}{\Lambda^4}\vec{\textbf{C}}^\textbf{T} {\boldsymbol{\sigma}}^{\left(2\right)} \vec{\boldsymbol{C}}.
\label{eq:eftsigmamatrix}
\end{equation}
\end{linenomath}
In order to find $\boldsymbol{\vec{\sigma}}^{\left(1\right)}$ and ${\boldsymbol{\sigma}}^{\left(2\right)}$, a system of linear equations has to be solved. It is obtained by substituting linearly-independent vectors $\vec{\boldsymbol{C}}$ into Eq.~\eqref{eq:eftsigmamatrix}. In the cross section calculation, the EFT interactions are implemented in the \textsc{FeynRules}~\cite{AguilarSaavedra:2018nen,Alloul:2013bka} package and interfaced with \MGvATNLO~\cite{Alwall:2014hca}. The NNPDF3.0LO~\cite{Ball:2014uwa} PDF set and $\alpS(M_{\PZ})=0.138$ were used in the calculation. In the EFT predictions, the SM contribution, $\sigma_{\tttt}^{\text{SM}}$ in Eqs.~\eqref{eq:eftsigma} and~\eqref{eq:eftsigmamatrix}, was rescaled to the NLO cross section of $9.2\unit{fb}$ for the collision energy of 13 \TeV.
The linear and quadratic coefficients, $\sigma_{k}^{\left(1\right)}$ and $\sigma_{{j},{k}}^{\left(2\right)}$, in Eq.~\eqref{eq:eftsigma} can be found in Table~\ref{tab:eftcoef}.

\begin{table}[!h]
	\topcaption{Linear (left) and quadratic (right) parameterization coefficients, $\sigma_{k}^{\left(1\right)}$ and $\sigma_{{j},{k}}^{\left(2\right)}$, of Eq.~\eqref{eq:eftsigma}. The coefficients $\sigma_{k}^{\left(1\right)}$ are in units (fb ${\TeVns}^{2}$), while the coefficients $\sigma_{{j},{k}}^{\left(2\right)}$ are in units (fb ${\TeVns}^{4}$).}
	\label{tab:eftcoef}
	\begin{center}
		\begin{tabular}{ l  c | c  c  c  c  c}
		                  & $\sigma_{k}^{\left(1\right)}$ &    &   &  $\sigma_{{j},{k}}^{\left(2\right)}$ & \\[3pt]
			Operator      &      & \WILOR  & \WILOLONE &  \WILOBONE & \WILOBEIGHT   \\[5pt]
			\hline
			\WILOR        & 0.39 &  5.59    & 0.36      & $-$0.39      & 0.3  \T \B\\
			\WILOLONE     & 0.47 &          & 5.49      & $-$0.45      & 0.13   \T \B\\
			\WILOBONE     & 0.03 &          &           & 1.9          & $-$0.08\T \B\\
			\WILOBEIGHT   & 0.28 &          &           &              & 0.45 \T \B\\
		\end{tabular}
	\end{center}
\end{table}

The observed limit of $\XSecEFTMuComboAllobs \sigmattttSM$,
with a corresponding expected limit of $\XSecEFTMuComboAllexp \sigmattttSM$ (assuming $\mu=1$), from the combined experimental
results, is used to constrain possible contributions of EFT operators.
Since the data are only sensitive to the ratios, $C_{k}/\Lambda^{2}$, the constraints are presented only for such ratios. In the limit setting, SM kinematics of the \tttt final state were assumed and only rate information was utilized to calculate the constraints. Besides the NLO scale uncertainty from the SM \tttt NLO prediction, no further scale uncertainties were added because other uncertainties on \tttt production are already included in the experimental limit.

Independent limits were obtained under the assumption that only one operator contributes to the~\tttt cross section with the coefficients of the other operators set to zero. The intervals obtained are summarized in Table~\ref{tab:independent}. More conservative estimates were obtained by marginalizing the contribution of other operators within the interval $C_{k}/\Lambda^{2} \in [-4\pi, 4\pi]$, defined by the stability of perturbation series. The corresponding limits are listed in Table~\ref{tab:marginal}. The results obtained are only slightly weaker than independent constraints because of the small correlations between the operators.

\begin{table}[!h]
    \topcaption{
      Expected and observed 95\% \CL intervals for selected coupling parameters. The intervals
      are extracted from upper limit on the \tttt production cross section in the EFT model,
      where only one selected operator has a nonvanishing contribution.}
      \label{tab:independent}
    \begin{center}
      \begin{tabular}{ l  c  c }
        Operator     & Expected $C_{k}/\Lambda^{2}$ (${\TeVns}^{-2}$) & Observed (${\TeVns}^{-2}$)\\
        \hline
	 \WILOR      & [$-$2.0, 1.8]    & [$-$2.1, 2.0] \T\B \\
	 \WILOLONE   & [$-$2.0, 1.8]    & [$-$2.2, 2.0] \T\B \\
	 \WILOBONE   & [$-$3.3, 3.2]    & [$-$3.5, 3.5] \T\B \\
	 \WILOBEIGHT & [$-$7.3, 6.1]    & [$-$7.9, 6.6] \T\B \\
       \end{tabular}
    \end{center}
  \end{table}

\begin{table}[!h]
    \topcaption{Expected and observed 95\% \CL intervals for selected coupling parameters when contribution of other operators is marginalized.}
    \label{tab:marginal}

    \begin{center}
      \begin{tabular}{ l  c  c }
        Operator     & Expected $C_{k}/\Lambda^{2}$ (${\TeVns}^{-2}$) & Observed (${\TeVns}^{-2}$)\\
        \hline
        \WILOR        & [$-$2.0, 1.9]     & [$-$2.2, 2.1] \T\B \\
        \WILOLONE     & [$-$2.0, 1.9]     & [$-$2.2, 2.0] \T\B \\
        \WILOBONE     & [$-$3.4, 3.3]     & [$-$3.7, 3.5] \T\B \\
        \WILOBEIGHT   & [$-$7.4, 6.3]     & [$-$8.0, 6.8] \T\B \\
      \end{tabular}
    \end{center}
  \end{table}

As shown in Tables~\ref{tab:independent} and \ref{tab:marginal}, the data have highest sensitivity to the contribution of \WILOR and \WILOLONE.
The allowed intervals for the coupling parameters are almost independent of the other considered operators and stay stable after marginalization.

\section{Summary}
\label{sec:conclusions}
A search for standard model \tttt production has been performed in
final states with one or two oppositely signed muons or electrons plus
jets.
The observed yields attributed to \tttt production are consistent with the background predictions. An upper limit at 95\% confidence level of \xsecfbcombo\unit{fb} is set on the cross section for \tttt production.
Combining this result with a previous same-sign dilepton and multilepton search~\cite{TOP-17-009} the resulting cross section is $\xsfbsscombounconstrained^{\,+\,\xsfbsscomboupunconstrained}_{\,-\,\xsfbsscombodownunconstrained}\unit{fb}$ with an observed significance of $\signiffbsscombo$ standard deviations.
The combined result constitutes
one of the most stringent constraints from CMS  on the production of
four top quarks and can be used for phenomenological reinterpretation
of a wide range of new physics models.
The experimental results are interpreted in the effective field theory
framework and yield limits on dimension-6 four-fermion operators
coupling to third generation quarks competitive with the latest ATLAS interpretation~\cite{ATLAStttt2018}.

\begin{acknowledgments}
We congratulate our colleagues in the CERN accelerator departments for the excellent performance of the LHC and thank the technical and administrative staffs at CERN and at other CMS institutes for their contributions to the success of the CMS effort. In addition, we gratefully acknowledge the computing centers and personnel of the Worldwide LHC Computing Grid for delivering so effectively the computing infrastructure essential to our analyses. Finally, we acknowledge the enduring support for the construction and operation of the LHC and the CMS detector provided by the following funding agencies: BMBWF and FWF (Austria); FNRS and FWO (Belgium); CNPq, CAPES, FAPERJ, FAPERGS, and FAPESP (Brazil); MES (Bulgaria); CERN; CAS, MoST, and NSFC (China); COLCIENCIAS (Colombia); MSES and CSF (Croatia); RPF (Cyprus); SENESCYT (Ecuador); MoER, ERC IUT, PUT and ERDF (Estonia); Academy of Finland, MEC, and HIP (Finland); CEA and CNRS/IN2P3 (France); BMBF, DFG, and HGF (Germany); GSRT (Greece); NKFIA (Hungary); DAE and DST (India); IPM (Iran); SFI (Ireland); INFN (Italy); MSIP and NRF (Republic of Korea); MES (Latvia); LAS (Lithuania); MOE and UM (Malaysia); BUAP, CINVESTAV, CONACYT, LNS, SEP, and UASLP-FAI (Mexico); MOS (Montenegro); MBIE (New Zealand); PAEC (Pakistan); MSHE and NSC (Poland); FCT (Portugal); JINR (Dubna); MON, RosAtom, RAS, RFBR, and NRC KI (Russia); MESTD (Serbia); SEIDI, CPAN, PCTI, and FEDER (Spain); MOSTR (Sri Lanka); Swiss Funding Agencies (Switzerland); MST (Taipei); ThEPCenter, IPST, STAR, and NSTDA (Thailand); TUBITAK and TAEK (Turkey); NASU and SFFR (Ukraine); STFC (United Kingdom); DOE and NSF (USA).

\hyphenation{Rachada-pisek} Individuals have received support from the Marie-Curie program and the European Research Council and Horizon 2020 Grant, contract Nos.\ 675440, 752730, and 765710 (European Union); the Leventis Foundation; the A.P.\ Sloan Foundation; the Alexander von Humboldt Foundation; the Belgian Federal Science Policy Office; the Fonds pour la Formation \`a la Recherche dans l'Industrie et dans l'Agriculture (FRIA-Belgium); the Agentschap voor Innovatie door Wetenschap en Technologie (IWT-Belgium); the F.R.S.-FNRS and FWO (Belgium) under the ``Excellence of Science -- EOS" -- be.h project n.\ 30820817; the Beijing Municipal Science \& Technology Commission, No. Z181100004218003; the Ministry of Education, Youth and Sports (MEYS) of the Czech Republic; the Lend\"ulet (``Momentum") Program and the J\'anos Bolyai Research Scholarship of the Hungarian Academy of Sciences, the New National Excellence Program \'UNKP, the NKFIA research grants 123842, 123959, 124845, 124850, 125105, 128713, 128786, and 129058 (Hungary); the Council of Science and Industrial Research, India; the HOMING PLUS program of the Foundation for Polish Science, cofinanced from European Union, Regional Development Fund, the Mobility Plus program of the Ministry of Science and Higher Education, the National Science Center (Poland), contracts Harmonia 2014/14/M/ST2/00428, Opus 2014/13/B/ST2/02543, 2014/15/B/ST2/03998, and 2015/19/B/ST2/02861, Sonata-bis 2012/07/E/ST2/01406; the National Priorities Research Program by Qatar National Research Fund; the Ministry of Science and Education, grant no. 3.2989.2017 (Russia); the Programa Estatal de Fomento de la Investigaci{\'o}n Cient{\'i}fica y T{\'e}cnica de Excelencia Mar\'{\i}a de Maeztu, grant MDM-2015-0509 and the Programa Severo Ochoa del Principado de Asturias; the Thalis and Aristeia programs cofinanced by EU-ESF and the Greek NSRF; the Rachadapisek Sompot Fund for Postdoctoral Fellowship, Chulalongkorn University and the Chulalongkorn Academic into Its 2nd Century Project Advancement Project (Thailand); the Welch Foundation, contract C-1845; and the Weston Havens Foundation (USA).
\end{acknowledgments}

\bibliography{auto_generated}

\cleardoublepage \appendix\section{The CMS Collaboration \label{app:collab}}\begin{sloppypar}\hyphenpenalty=5000\widowpenalty=500\clubpenalty=5000\vskip\cmsinstskip
\textbf{Yerevan Physics Institute, Yerevan, Armenia}\\*[0pt]
A.M.~Sirunyan$^{\textrm{\dag}}$, A.~Tumasyan
\vskip\cmsinstskip
\textbf{Institut f\"{u}r Hochenergiephysik, Wien, Austria}\\*[0pt]
W.~Adam, F.~Ambrogi, T.~Bergauer, J.~Brandstetter, M.~Dragicevic, J.~Er\"{o}, A.~Escalante~Del~Valle, M.~Flechl, R.~Fr\"{u}hwirth\cmsAuthorMark{1}, M.~Jeitler\cmsAuthorMark{1}, N.~Krammer, I.~Kr\"{a}tschmer, D.~Liko, T.~Madlener, I.~Mikulec, N.~Rad, J.~Schieck\cmsAuthorMark{1}, R.~Sch\"{o}fbeck, M.~Spanring, D.~Spitzbart, W.~Waltenberger, C.-E.~Wulz\cmsAuthorMark{1}, M.~Zarucki
% ...
\vskip\cmsinstskip
\dag: Deceased\\
1:  Also at Vienna University of Technology, Vienna, Austria\\
% ...
\end{sloppypar}
\end{document}